\documentstyle[psfig,graphicx,epsf,amssymb]{mn} 
\title[Cold gas and star formation in a merging galaxy sequence] {Cold gas and
star formation in a merging galaxy sequence} 

\author[Georgakakis, Forbes \&  Norris] 
{Antonis Georgakakis$^{1}$\thanks{\sf age@star.sr.bham.ac.uk}, 
  Duncan A. Forbes$^{1,2}$\thanks{\sf forbes@star.sr.bham.ac.uk}, 
  Ray P. Norris$^{3}$\thanks{\sf rnorris@atnf.csiro.au} \\ \\ 
  $^1$ School of Physics and Astronomy, University of Birmingham,
  Edgbaston, Birmingham, B15 2TT, UK\\
  $^2$ Astrophysics \& Supercomputing, Swinburne University,
  Hawthorn, VIC 3122, Australia\\
   $^3$ Australia Telescope National Facility, CSIRO, Epping,
NSW, Australia\\ 
}

\begin{document}
\maketitle  

\begin{abstract}
We explore the evolution of the cold gas (molecular and neutral hydrogen)
and star-formation activity during galaxy interactions, using a merging
galaxy sequence comprising both pre- and post-merger candidates. Data for
this study come from the literature but supplemented by some new radio   
observations presented here. Firstly, we confirm that the ratio of
far-infrared luminosity to molecular hydrogen mass ($L_{FIR}/M(H_{2})$;
star-formation efficiency) increases close to nuclear coalescence. After
the merging of the two nuclei there is evidence that the star-formation
efficiency declines again to values typical of ellipticals. This trend 
can be attributed to $M(H_{2})$ depletion due to interaction induced 
star-formation. However, there is significant scatter, likely to arise from
differences in the interaction details (e.g. disk-to-bulge ratio, geometry)
of individual systems.  Secondly, we find that the central molecular
hydrogen surface density, $\Sigma_{H_{2}}$, increases close to the
final stages of the merging of the two nuclei. Such a trend, indicating
gas inflows due to gravitational instabilities during the interaction, is
also predicted by numerical simulations.   Furthermore, there is evidence
for a decreasing fraction of cold gas mass from early interacting systems
to merger remnants, attributed to neutral hydrogen conversion into other
forms (e.g. stars, hot gas) and molecular hydrogen depletion due to
on-going star-formation.  The evolution of the total-radio to blue-band
luminosity ratio, reflecting the total (disk and nucleus) star-formation
activity, is  also investigated. Although this ratio is on average higher
than that of isolated spirals, we find a marginal increase along the
merging sequence, attributed to the relative insensitivity of disk
star-formation to interactions. However, a similar result is also obtained
for the nuclear radio emission, although galaxy interactions are believed
to significantly affect the activity (star-formation, AGN) in the central
galaxy regions. Nevertheless, the nuclear-radio to blue-band luminosity
ratio is significantly elevated compared to isolated spirals. Finally, we
find that the FIR--radio flux ratio distribution of interacting galaxies is 
consistent with star-formation being the main energising source. 
\end{abstract} 
 

\section{Introduction}\label{intro}

Tidal interactions and mergers are believed to play a significant role in
the evolution of galaxies. Such phenomena can not only enhance the activity
in and around the  nuclear region (star-formation or AGN) but can also
irreversibly alter the morphological appearance of the participant galaxies.
Toomre \& Toomre (1972) were the first to demonstrate that gravitational
interactions can give rise to tidal features (e.g. bridges, tails) and also
proposed the merging of disk galaxies as a plausible formation scenario for
ellipticals (called the merger hypothesis). Indeed, recent more sophisticated
numerical simulations have shown that dynamical friction and violent
relaxation during disk-galaxy interactions will disrupt any pre-existing
disks leading to relaxed $r^{1/4}$--law light profiles similar to those of
ellipticals (Barnes 1988, 1992; Hernquist 1992, 1993). The same gravitational
instabilities can produce significant gas inflows towards the centre of the
galaxy, where  enhanced star-formation activity is likely to take place
(e.g. Mihos \& Hernquist 1996).  

Indeed, high molecular gas densities have been observed in the central 
regions of the IRAS starburst galaxies, thought to be gas rich systems
close to the final stages of merging (Kennicutt 1998; Planesas et
al. 1997; Sanders \& Mirabel 1996). The high molecular gas density regions
are also found to be associated with enhanced nuclear star-formation
(and/or AGN)  activity as inferred from their far-infrared (FIR; Kennicutt
1998), radio  (Hummel 1981; Hummel et al. 1990) and  optical emission-line
luminosity  (Keel et al. 1985).   A smaller but systematic enhancement
compared to isolated spirals is also seen in the disk radio power (Hummel
1981) and disk H$\alpha$ emission (Kennicutt et al. 1987) which is again 
attributed to interaction induced star-formation activity. Additionally,
the fraction of interacting systems found in IRAS-selected samples
increases with FIR  luminosity (Lawrence 1989; Gallimore \& Keel  1993),
suggesting that collisions play a major role in triggering powerful
starbursts.  

Evidence also exists linking merger remnants with elliptical galaxies. 
For example,  merger remnants tend to have optical and/or near--infrared
light profiles that follow the $r^{1/4}$--law (Joseph \& Wright
1985). Secondly,  many, otherwise `normal' ellipticals, exhibit low surface
brightness loops and shells (Malin \& Carter 1980; Schweizer \& Seitzer
1988) that are likely to be due to past disk-galaxy
encounters. Recently, Forbes, Ponman \& Brown (1998) showed that late stage
disk-disk mergers and ellipticals with young stellar populations deviate
from the fundamental plane of ellipticals. This can be understood in terms
of a centrally located starburst induced by a gaseous merger event.  

The significance of tidal interactions in the evolution of galaxies has
motivated a number of studies aiming to explore the properties of
interacting systems at different stages during the encounter. 
Toomre (1977) first proposed a merging sequence of eleven peculiar galaxies
spanning a range of post- to pre-mergers (the `Toomre sequence') and
suggested that the final product of the interaction  is likely to resemble
an elliptical galaxy. Keel \& Wu (1995) used morphological criteria to
define a merging galaxy sequence by  assigning  a merger stage number to
each galaxy pair or merger remnant. They found that indicators of on-going 
star-formation activity, such as the $U-B$, $B-V$ colours and the
FIR--to--blue-band luminosity ratio tend to peak close to the final stages
of nuclear coalescence and then  decrease at post-merger stages to attain
values typical of ellipticals.  A similar result was obtained by Casoli et
al. (1991) who also studied the evolution of star-formation activity
estimators (FIR  temperature, FIR--to--blue-band luminosity, FIR luminosity
to molecular hydrogen mass) for a small merging galaxy sequence defined by
morphological criteria. More recently, Gao \& Solomon (1999) used the
projected separation between the nuclei of merging FIR-selected galaxies as
an estimator of the interaction stage. They found clear evidence for
increasing star-formation efficiency (SFE; estimated by the ratio of
FIR-luminosity to molecular hydrogen mass)  with decreasing nuclei
separation. They argue that this is primary due to the depletion of the
molecular gas reservoirs of these systems by on-going star-formation
triggered by interactions. Hibbard \& van Gorkom (1996) studied the cold
gas properties and the dynamics of a small sample of pre- and post-mergers
from the Toomre sequence. They find striking differences in the
distribution of H\,I in pre- and post-mergers, with increasing fractions of
H\,I outside the optical bodies at later stages. They argue that during the
interaction about half of the cold gas material is ejected in tidal
features, whereas the atomic gas remaining within the original disks is
either converted into stars or heated up to X-ray temperatures.   
Read \& Ponman (1998) investigated the X-ray evolution of a similarly
defined merging sequence. Although  they also found a rise and fall in the
X-ray--to--blue-band luminosity around nuclear coalescence, the increase is
by a factor of ten smaller than that seen in the FIR--to--blue-band 
luminosity. They argue that this is likely to be due to superwinds blowing
out the hot X-ray emitting gas.  These studies clearly indicate that large
changes occur in the energetic, structural and kinematic properties of
galaxies during interactions and mergers.

The above mentioned  studies either concentrated  only on
pre-mergers (e.g. Gao \& Solomon 1999) or investigated the properties of
small samples of pre- and post-merger galaxies (assumed to be
representative), albeit in great detail (Hibbard \& van Gorkom 1996; Casoli
et al. 1991;  Read \& Ponman 1998). In this paper we have compiled a
large sample of interacting systems from the literature spanning a wide
range of pre- and  post-merger stages, aiming to explore the evolution of
both their star-formation and their cold gas (molecular and neutral
hydrogen) properties.  Additionally, comparison of the galaxy properties
along the merger sequence with those of `normal' ellipticals and isolated
spirals allows investigation of the merging hypothesis for the formation of
ellipticals.   

In section \ref{sample} we discuss the sample selection, while section
\ref{observations} describes the new radio observations carried out for a
selected sample of interacting galaxies. Section \ref{results} presents the
results from our analysis. Finally, in section \ref{conclusions} we
summarise our conclusions. Throughout this paper we assume a value
$H_{o}=75\,\mathrm{km\,s^{-1}\,Mpc^{-1}}$.

\section{The sample}\label{sample}

Compiling a sample of interacting galaxies and merger remnants from the
literature is problematic. Different authors have used different selection
criteria (e.g. FIR, morphological selection) that are likely to 
introduce biases against certain types of interactions. In this study,
we merged several interacting galaxy samples from the literature, with
different selection criteria in an attempt to minimise any selection
biases. However, it should be noted that most of the galaxies in this study
are FIR luminous and are also biased against mergers occurring along our
line of sight. Therefore, although the present sample is by no means
statistically 
complete, it could  be regarded as representative of interacting systems
and merger remnants spanning a wide range of properties.  
Our sample is largely culled from the following studies:   

\begin{enumerate}

\item Keel \& Wu (1995) selected nearby pre- and post-merger 
galaxies based on their optical  morphology  and ordered them into a
sequence  by assigning a dynamical `stage' number to each galaxy pair or 
merger remnant.  

\item Gao \& Solomon (1999) and Gao et al. (1998) 
compiled samples of FIR-luminous and ultra-luminous galaxies with available
CO(1-0) observations (providing an estimate of the available molecular
hydrogen mass, $M(H_2)$).  These samples consist exclusively of
pre-mergers.  

\item Surace et al. (1993) presented a sample of merging galaxies,
morphologically selected from the $60\mu m$ flux density 
limited IRAS Bright Galaxy Sample (Soifer et al. 1987). 

\end{enumerate}

We focus on interacting systems and merger remnants from the above mentioned
samples satisfying the following criteria: 

{\bf (a)} $\delta r\lesssim (D_{1}+D_{2})/2$, where
$\delta r$ is the separation between the two nuclei of the merging galaxies
and $D_1$, $D_2$ their major axis diameters. This selection criterion is
similar to that employed by Gao \& Solomon (1999).

{\bf (b)} recessional velocities $v\lesssim13\,000\,\mathrm{km\,s^{-1}}$, 
corresponding to a distance $\lesssim170$\,Mpc.

{\bf (c)}  far-infrared luminosities $L_{FIR}<10^{12}\,L_{\odot}$.

{\bf (d)}  we only consider disk mergers by discarding pairs for which
there is morphological evidence that at least one of the components is
elliptical.   

{\bf (e)} we attempt to restrict our sample to `major mergers';
i.e. mergers involving galaxies of similar mass, in view of their relevance
to the formation of elliptical galaxies. Therefore, we only consider pairs
in which the individual components have a $B$-band magnitude difference of
less than 1.5\,mag, corresponding to a mass ratio 1:4 (assuming the same
mass-to-light ratio). Mergers involving higher mass ratios may merely
puff-up the disk and/or perhaps enlarge the bulge, but will not completely
rearrange the light profile of the galaxy.
However, the $B$-band magnitudes are affected by star-formation and
do not provide a sensitive estimator of the total mass of a
system. Nevertheless, to the first approximation they should provide a
rough  estimate of the galaxy mass ratio that is sufficient for the
purposes of this paper.   

The sample employed in this study is presented in Table 1, which has the
following format

{\bf  1.} galaxy names.

{\bf  2.} heliocentric distance, $D$, in Mpc, assuming $H_{o}= 75 \,
\mathrm{ km \, s^{-1} \, Mpc^{-1}}$.  No correction for the Local Group 
velocity or the Virgocentric flow has been applied. However, these
corrections are not expected to modify the estimated distances by more
than $10\%$. Moreover, in our analysis we consider ratios of observed
quantities that are independent of distance. 

{\bf 3.} total radio flux density at 1.4\,GHz (20\,cm;
$S^{tot}_{1.4}$) in mJy. For most galaxies in the present sample
$S^{tot}_{1.4}$ was obtained from Condon et al. (1991) and from the NRAO
VLA Sky Survey (NVSS) catalogue (Condon et al. 1998).

{\bf 4.} galaxy `age' parameter. 
Each galaxy is assigned  an `age' parameter, relative to the time of the
merging of the two nuclei. Negative `ages' are for pre-mergers  while
positive `ages' correspond to merger remnants. For pre-mergers the 
`age' is estimated by dividing the projected separation of the two nuclei,
$\delta r$, by an (arbitrary) orbital decay velocity
$v=30\,\mathrm{km\,s^{-1}}$.  It is clear that the `age' parameter for 
pre-mergers is affected by projection effects or different interaction
geometries. However, to the first approximation, it provides an estimate of
the stage of the merging and allows plotting of pre- and post-mergers on
the same scale.
For post-mergers we adopt the evolutionary sequence defined by  Keel \& Wu
(1995) using dynamical and morphological criteria. In particular, the `age'
parameter for these systems is calculated by multiplying the dynamical
stage number, defined by  Keel \& Wu (1995), by the factor
$4\times10^{8}$\,yr. This conversion factor is found to be 
appropriate for  the 3 merger remnants in the Keel \& Wu sample with
available spectroscopic estimates (i.e. NGC 2865, NGC 3921, NGC 7252;
Forbes, Ponman \& Brown  1998). It should be stressed that the `age'
parameter for both pre- and post-mergers does not represent an absolute
galaxy age but is an indicator of the evolutionary stage of the
interaction.

{\bf 5.} far-infrared luminosity in solar units
($L_{\odot}=3.83\times10^{26}$\,W)

\begin{equation}\label{eq_lfir}
L_{FIR}=4\pi\,D^{2}\times 1.4 \times S_{FIR},
\end{equation}

\noindent where $S_{FIR}$ is the FIR flux in $\mathrm{W\,m^{-2}}$
between $42.5$ and $122.5\mu m$ (Sanders \& Mirabel 1996)

\begin{equation}\label{eq_fir} 
S_{FIR}(\mathrm{W\,m^{-2}})=1.26\times\,10^{-14}\times(2.58\times f_{60} + f_{100}), 
\end{equation}

\noindent where $f_{60}$ and $f_{100}$ are the IRAS fluxes at 60 and
100$\mu m$  respectively in Jansky. The scale factor 1.4 in equation
(\ref{eq_lfir}) is the correction factor required to account principally
for the extrapolated flux longward of the IRAS 100$\mu$m filter  (Sanders
\& Mirabel 1996).    

{\bf 6.} molecular hydrogen mass, $M(H_2)$, estimated from the
CO(1-0) emission. The sources from which the CO(1-0) intensity measurements 
were obtained are given in Table 1. The conversion factor
$N(H_2)/I_{CO}=3\times10^{20} \,  \mathrm{cm^{-2}\,(K\,km\,s^{-1})^{-1}}$,
appropriate for molecular clouds in the Milky Way (Sanders, Solomon \&
Scoville 1984) was adopted. It should be noted that use of this conversion
factor assumes  that the mean properties of the molecular gas in distant
galaxies (i.e. density, temperature and metallicity) are similar to those 
of the Milky Way. 
However, the molecular clouds of the interacting systems studied here 
are likely to have both higher densities and temperatures and different
metallicities compared to those of the Milky  Way. These effects are
expected to modify the CO-to-$M(H_2)$ conversion  factor for these
galaxies. 
Indeed, a number of studies suggest that use of the standard  Galactic 
conversion factor for starbursts overestimates their $M(H_2)$ (Maloney \&
Black 1988; Tinney et al. 1990; Solomon  et al. 1997) and therefore a
smaller  CO-to-$M(H_2)$ factor is appropriate for these galaxies.
Nevertheless, to facilitate comparison of  our results with other  
studies we use the standard Galactic $N(H_2)/I_{CO}$ conversion factor.  
In any case the results can be interpreted in terms of CO luminosity
($L_{CO}$) rather than $M(H_2)$, since a constant scaling factor is used
throughout. 

{\bf 7.} neutral hydrogen mass, $M(HI)$. The H\,I masses are
related to the H\,I integrated intensities, $F(H\,I)$ (measured in 
$\mathrm{Jy\,km\,s^{-1}}$), by

\begin{equation}
M_{H\,I} (M_{\odot})=2.356\times10^{5}\,\times F(H\,I)\,\times (D/\mathrm{Mpc})^{2}.
\end{equation}

\noindent The sources from which the H\,I intensity measurements were
obtained are also given in Table 1. 

{\bf 8.} total $B$-band magnitude, $B_T$. This has been corrected for
Galactic extinction but not for internal extinction. This is because the
systems studied here have disturbed morphologies and are likely to have
more dust than normal spirals. Therefore, any correction for internal
extinction which is based on the galaxy morphology (like that introduced in
the RC3 catalogue by de Vaucouleurs et al. 1991) is expected to be
unreliable. In few cases, RC3 did not provide total $B$-band magnitudes and
instead we used the total magnitudes from the catalogue compiled by Garnier
et al. (1996).  

{\bf 9.} Central surface density of molecular hydrogen, $\Sigma_{H_{2}}$,
in units of $\mathrm{M_{\odot}\,pc^{-2}}$. This is calculated from high
resolution CO(1-0) observations by dividing the (unresolved) flux within
the synthesised beam by its area (in pc$^2$). A consequence of this
definition is an increase of the linear size of the region enclosed by the 
synthesised beam with distance. However, to the first approximation,
$\Sigma_{H_{2}}$ is representative of the molecular gas concentration of
different systems.  The $\Sigma_{H_{2}}$ for interacting systems and merger
remnant candidates were mainly obtained from Kennicutt (1998) and Planesas
et al. (1997).

{\bf 10-12.} Central radio flux density at 1.49\,GHz (20\,cm),
4.79\,GHz (6\,cm) and 8.44\,GHz (3\,cm) respectively, integrated within 
an aperture of $\approx2$\,kpc diameter at the distance of the galaxy. We
use high resolution radio maps available on NED (1.49\,GHz: Condon et
al. 1990;  8.44\,GHz: Condon et al. 1991) as well as our own radio observations
at 4.79 and 8.64\,GHz (see section \ref{observations}). Upper limits are
$3\sigma$ estimates, where $\sigma$ is the RMS noise within a beam. 

{\bf 13.} Radio spectral index, $\alpha$, derived from the central 
$2$\,kpc diameter aperture flux densities using the relation
\begin{equation}
\alpha=\mathrm{\log(S_{\nu_{1}}/S_{\nu_{2}})/\log(\nu_{1}/\nu_{2})}
\end{equation}

\begin{figure*}
\centering
\vspace{-2.5cm}
\hspace*{-1cm}
\includegraphics[width=1.2\textwidth, height=1.2\textheight,angle=-180]{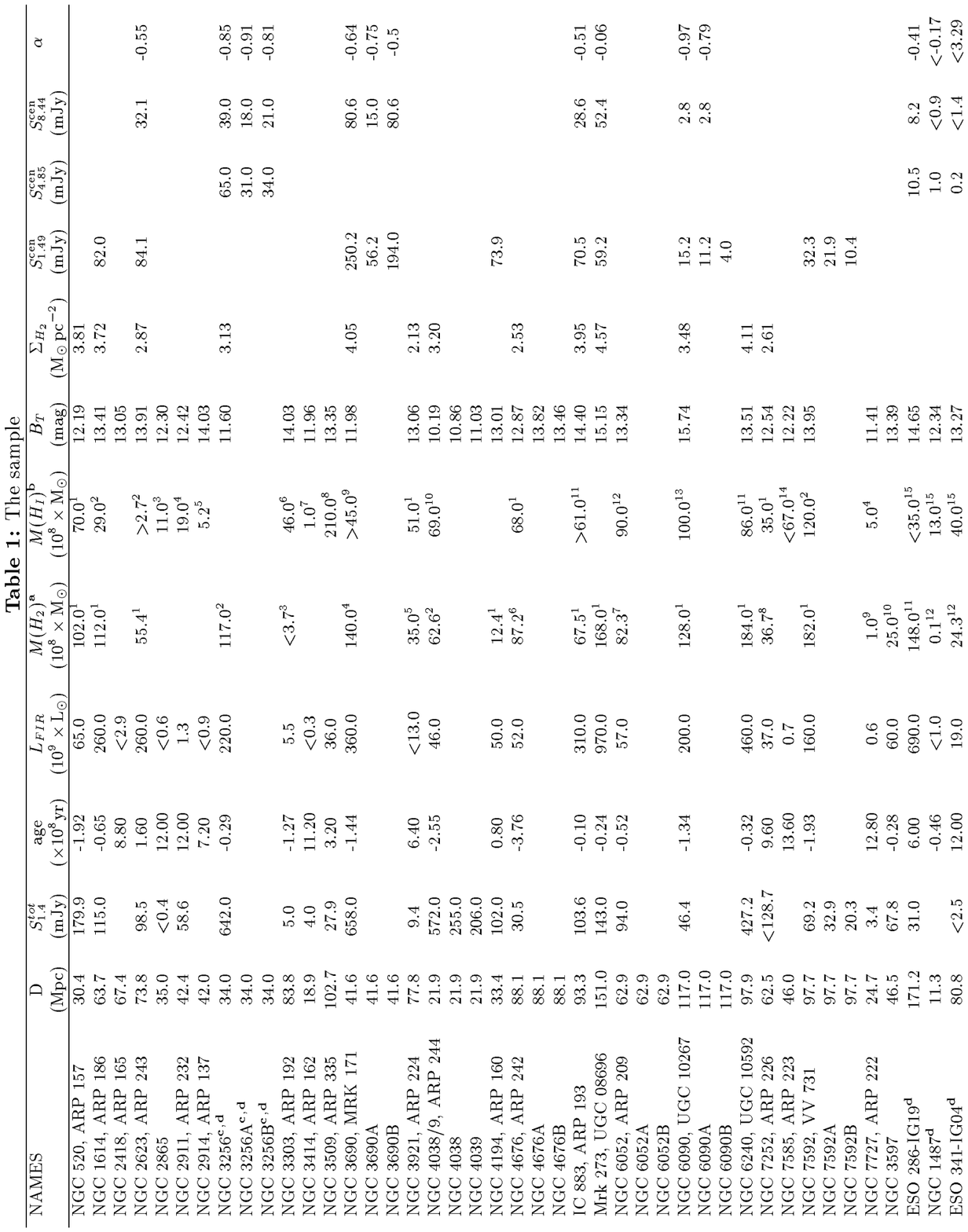}
\end{figure*}

\begin{figure*}
\centering
\vspace{-2.5cm}
\hspace*{-1cm}
\includegraphics[width=1.2\textwidth, height=1.2\textheight,angle=-180]{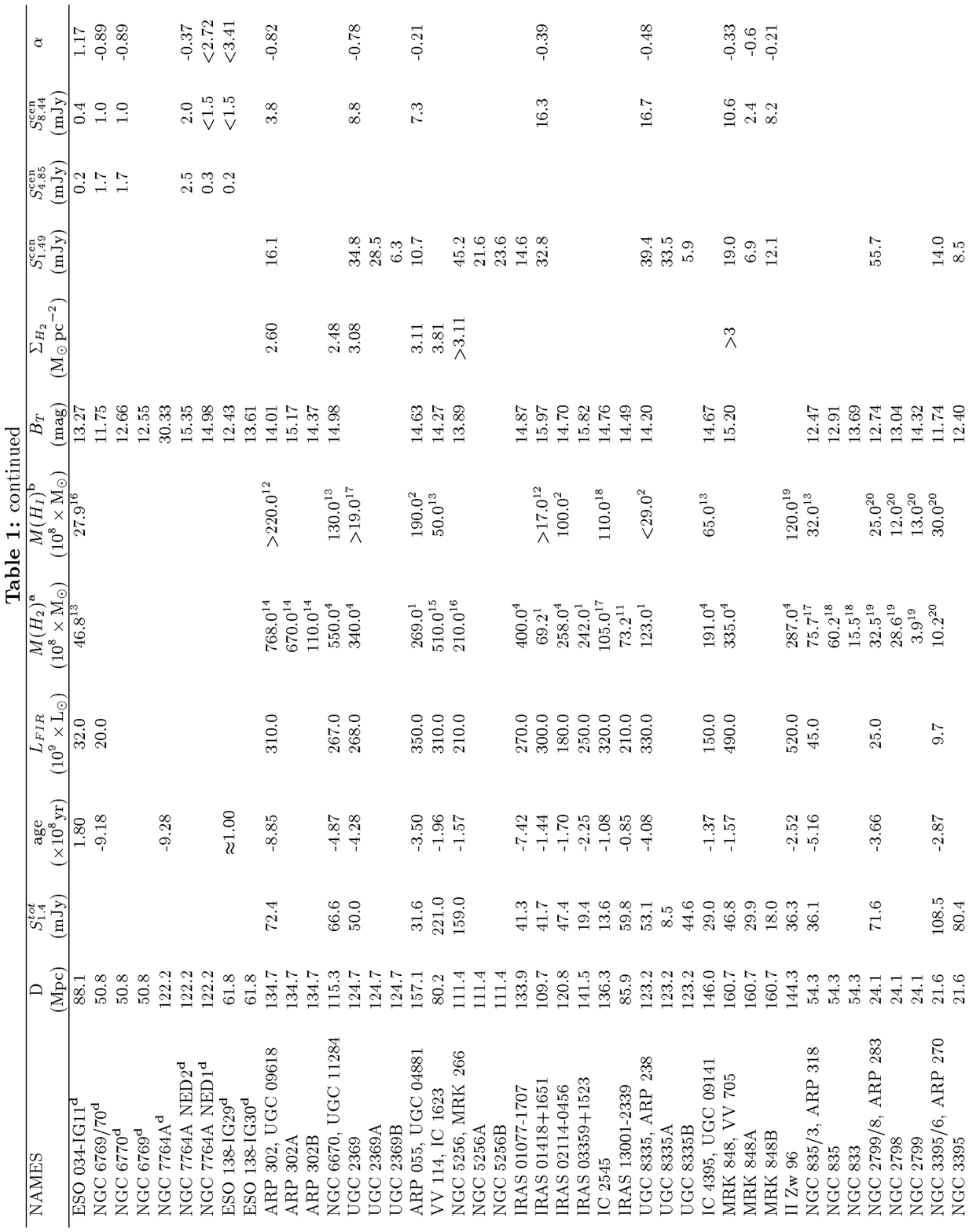}
\end{figure*}

\begin{figure*}
\centering
\vspace{-2.5cm}
\hspace*{-1cm}
\includegraphics[width=1.2\textwidth, height=1.2\textheight,angle=-180]{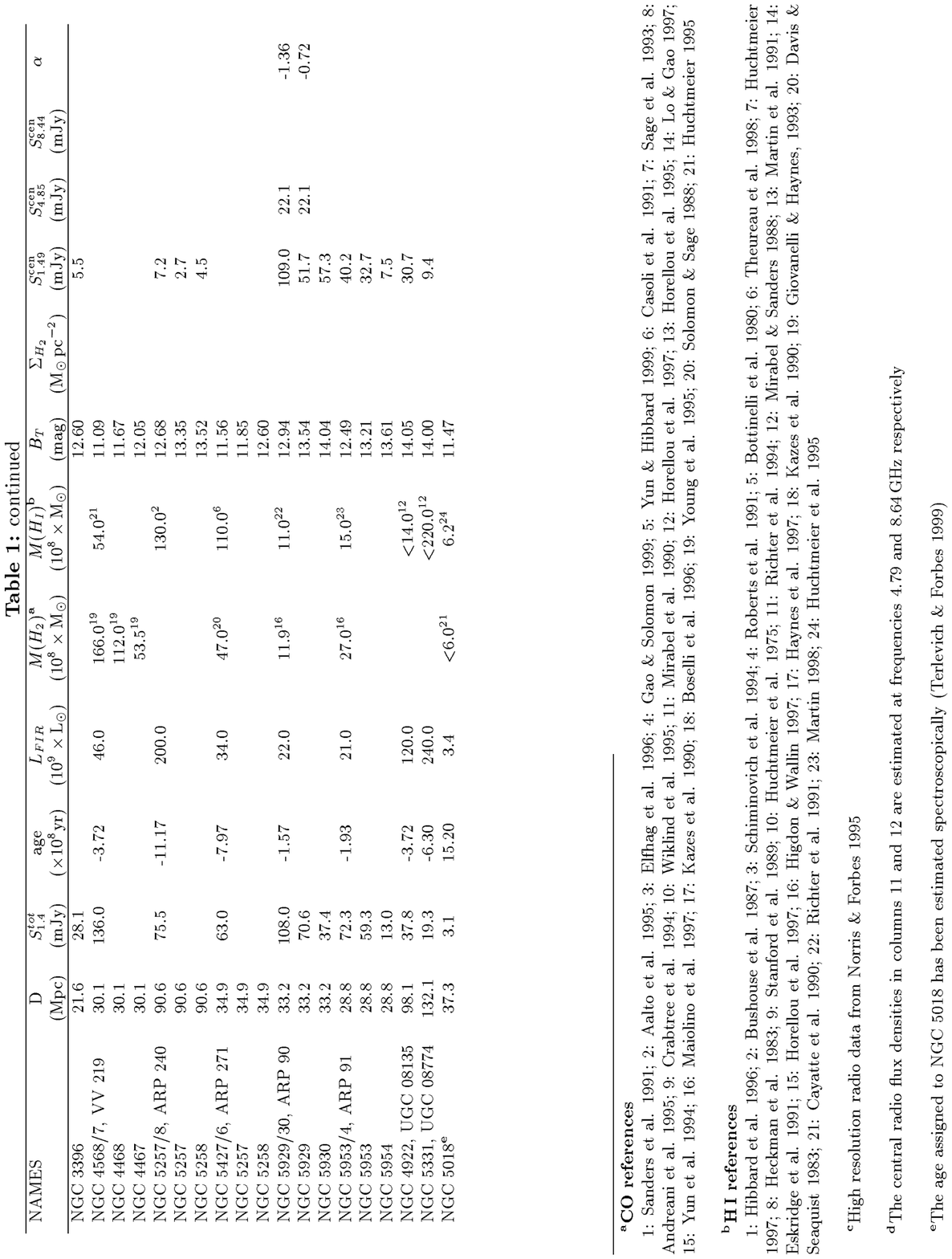}
\end{figure*}

\section{New radio observations}\label{observations}

To explore the effect of gravitational encounters to the nuclear radio
emission of on-going interacting systems and merger remnants we have
complemented high resolution radio data in the literature with our own 
new radio observations of seven systems (NGC 6769/70, ESO 286-IG19,
NGC 1487, ESO 034-IG11, NGC 7764A, ESO 138-IG29, ESO
341-IG04). Also two of these systems (ESO 034-IG11 and ESO
138-IG29) are ring galaxies, allowing us to comment on the radio
properties of this  
class of encounters. Interactions that produce a ring system  are 
somewhat different to the disk-galaxy interactions/mergers studied here. In
this case the intruder galaxy makes a rapid near perpendicular approach to
the disk of the primary galaxy. Unlike more planar interactions, the disk
is little affected until the intruder passes through it. Such collisions 
produce merger remnants that are different from  those resulting from
disk-galaxy encounters and therefore we consider them separately from the
main sample. In all the figures,  ESO 034-IG11 and ESO 138-IG29 systems are
plotted as post-mergers, although we differentiate them from the rest of
the merger remnants with different symbols.

The new radio observations were carried out using the Australia
Telescope Compact Array at 4.79\,GHz (6\,cm) and 8.64\,GHz (3\,cm)
simultaneously. We  alternated observations of the target galaxy and a
phase calibrator throughout the observing run. Our observational
parameters are given in Table  2. The same amplitude calibrator
(1934--638) was observed with each galaxy, with  an assumed flux 
density of 5.83\,Jy at 4.79\,GHz and 2.84\,Jy at 8.64\,GHz. The data
were edited, calibrated and CLEANed using the {\sc aips}  software
package. The typical half--power beam--width (HPBW) of the final
images are $\approx2$\,arcsec at 4.79\,GHz and $\approx1$\,arcsec at
8.64\,GHz.      

Because these observations and their analysis were optimised for studying
the nuclear region, they are relatively insensitive to extended emission.
We will therefore concentrate on the emission within the central
10\,arcsec.  Details on individual galaxies are given in Appendix $A$. Also
shown in Appendix $A$ are the radio contours at 4.79\,GHz
overlayed\footnote{These images were produced using the {\sc kview}
application from the {\sc karma} software suite (Gooch 1995)} on the
optical images (from Digital Sky Survey) of the seven galaxies in
Table 2 (Figures A1-A7). 

\setcounter{table}{1}

\begin{table*} 
\footnotesize 
\begin{center} 
\begin{tabular}{l c c c c c c c c}
\multicolumn{9}{l}{{\bf Table 2.} Observational parameters} \\ \hline
Galaxy & Obs. & Phase      & 3cm  & 3cm  & 3cm    & 6cm  & 6cm   & 6cm \\  
       & Date & Calibrator & Beam & P.A. & Noise  & Beam & P.A.  & Noise \\
 &     &               &(arcsec$^2$) & (deg)  & ($\mu$Jy) &(arcsec$^2$) &
(deg) & ($\mu$Jy) \\  \hline

NGC 6769/70 & 1995 Dec. 10 & 1925--610
& 1.08 $\times$ 0.96 & 29.0 & 30 
& 2.02 $\times$ 1.80 & 31.0 & 30\\
ESO286--IG19 & 1995 Dec. 10 & 2058--425
& 1.33 $\times$ 1.03 & 13.4 & 50
& 2.47 $\times$ 1.87 & 9.8 & 40\\
NGC 1487 & 1995 Dec. 11 & 0355--483 
& 1.48 $\times$ 0.93 & --15.1 & 30
& 2.84 $\times$ 1.71 & --16.3 & 30\\
ESO034--IG011 &1995 Dec. 12 & 0757--737
& 1.31 $\times$ 0.87 & --66.5 & 30 
& 2.37 $\times$ 1.55 & --73.4 & 30\\
NGC 7764A & 1998 Jan. 28 & 0008--421 
& 1.80 $\times$ 1.17 & 18.9 & 50 
& 3.25 $\times$ 2.11 & 18.0 & 42\\
ESO138--IG29 & 1998 Jan. 24& 1718--649 
& 1.44 $\times$ 1.10 & 87.8 & 50 
& 2.56 $\times$ 1.99 & 87.1 & 46\\
ESO341--IG04 & 1998 Jan. 10& 2106--413 
& 1.84 $\times$ 1.27 & 27.4 & 47 
& 3.38 $\times$ 2.22 & 23.4 & 37\\ \hline
\end{tabular}
\end{center} 
\end{table*}

\section{Results and discussion}\label{results}

This section studies the evolution of estimators of the star-formation
activity and the cold gas content of interacting systems as a function
of the `age' parameter. In particular, the significance of any
correlations between these quantities, including upper limits, is
investigated  using the Spearman rank correlation analysis  
implemented in the {\sc asurv} package (LaValley, Isobe  \& Feigelson 1992;
Isobe, Feigelson \& Nelson 1986). Throughout this paper we assume that
the independent parameter in the Spearman rank correlation test is
`age'.

\subsection{Star-formation efficiency}

The ratio  $L_{FIR}/M(H_2)$ estimates the number of massive stars formed 
per molecular cloud and is thus, related to the integrated galaxy
star-formation efficiency (SFE).   
The molecular gas mass is estimated from the galaxy CO(1-0) emission using
the standard CO--to--$M(H_{2})$ conversion factor (section
\ref{sample}) to facilitate comparison of our results with other
studies. It should be noted however that the CO--to--$M(H_{2})$
conversion  factor is likely to vary, depending on the  galaxy physical
conditions (see discussion below).  Therefore, the CO(1-0) emission is
strictly estimating the CO mass rather than $M(H_{2})$.   

\begin{figure} 
\centerline{\psfig{figure=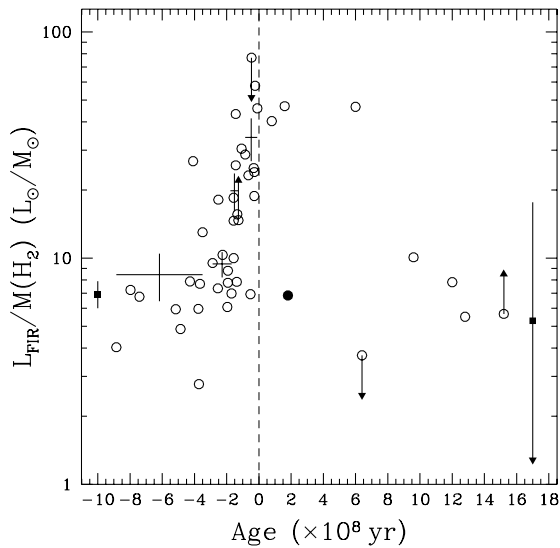,width=0.45\textwidth,angle=0}} 
\caption{ Star-formation efficiency ($SFE=L_{FIR}/M(H_2)$) as a function
of the galaxy `age' parameter. Negative `ages' are for pre-mergers while
positive `ages' correspond to post-mergers. The dashed line signifies the
time of nuclear coalescence (`age'=0) and separates pre- and post-merger
systems. Open circles are the galaxies in the present sample. The filled
circle represents the ring galaxy ESO 034-IG11. Filled 
squares correspond to the mean $SFE$ for (i) isolated spirals (left;
Solomon \& Sage 1988) and (ii) ellipticals (right; Lees et
al. 1991).  The crosses signify the mean $SFE$ for pre-mergers averaged
within `age'-parameter bins. The horizontal  error bars represent the width
of each bin, selected so that each bin comprises about 10 systems. The
vertical error bars are the standard error on the mean SFE within each bin.
There is evidence that the SFE starts at a level similar to that of
isolated spirals, peaks at nuclear coalescence and then declines at
post-merger stages to a level similar to that of normal ellipticals. 
}\label{fig_sfe} 
\end{figure}

Figure \ref{fig_sfe} plots the SFE as a function of the galaxy `age'
parameter. As explained in section \ref{sample} negative `ages' correspond
to pre-merger stages, zero corresponds to nuclear coalescence, while 
positive `ages' are for merger remnants.  It is clear from Figure
\ref{fig_sfe} that there is a trend of increasing SFE as the interaction
progresses  towards the final stages of nuclear coalescence. 
Indeed, we calculate a Spearman rank
correlation coefficient $r=0.68$, corresponding to a probability that
the correlation arises by chance  $P<0.01\%$. 
At later stages, the situation is less clear  due to the small number of  
systems with available CO measurements. Nevertheless, there is evidence
that throughout the merger process the SFE starts at a level comparable to
isolated spirals, peaks around nuclear coalescence and decreases at
post-mergers to a level similar to that of normal ellipticals. Similarly, 
Gao \& Solomon (1999) found an increase in the mean SFE of interacting
galaxy pairs with decreasing nuclear separation.  Moreover, Solomon \& Sage
(1988) found that strongly interacting and merging galaxies (i.e. galaxies
exhibiting tidal features such as tails and bridges)  have SFEs that are
about an order of magnitude higher than that of isolated and weakly
interacting galaxies.   

It is also clear from Figure \ref{fig_sfe} that there is significant
scatter in the SFE evolution of pre-mergers. This is likely to be partly
due to projection effects. We attempt to  compensate for these effects by
averaging the data for pre-mergers in different `age' bins.  The  bins have
variable widths so that they all  comprise similar number of points
($\approx10$). The mean SFE and the standard error is then  calculated
within each bin. The results are also shown in Figure \ref{fig_sfe}. It is
clear that the mean SFE increases toward the final coalescence of the two
nuclei. 

\begin{figure} 
\centerline{\psfig{figure=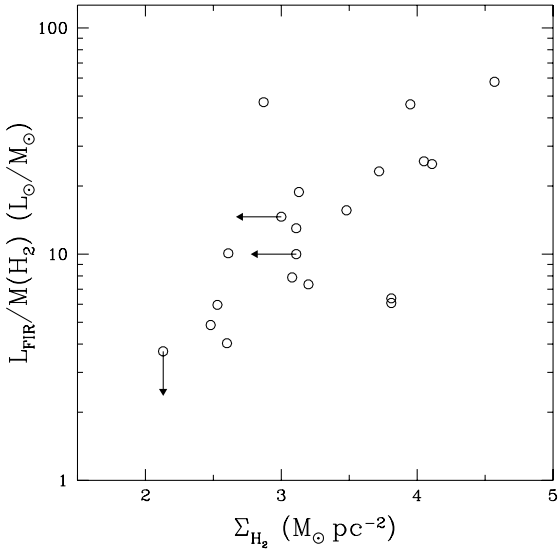,width=0.45\textwidth,angle=0}} 
\caption{ Star-formation efficiency as a function of central molecular gas 
surface density.  There is evidence that the SFE and $\Sigma_{H_{2}}$ are
correlated, indicating that high central concentrations of molecular
hydrogen are also associated with powerful starbursts.
}\label{fig_sh2sfe} 
\end{figure}

However, the observed scatter in the SFE evolution of pre-mergers is also
expected to be due to differences in the details of individual interactions
(e.g. geometry, initial conditions, bulge-to-disk ratio). Mihos \&
Hernquist (1996) carried out numerical simulations to explore the effect of
galaxy structure on the gas dynamics and evolution of starburst activity in
mergers. They found that in the case of galaxies with dense central bulges
significant gas inflows occur close to the final stages of merging. On the
contrary, gas inflows and thus, the peak of star-formation in bulgless
galaxies occur earlier in the interaction. As a result some of  the gas in
these systems is depleted at early stages and only a relatively weak
starburst is expected during nuclear coalescence. 
As demonstrated in Figure \ref{fig_sh2sfe}, where we plot SFE against
$\Sigma_{H_2}$, gas inflows and the resulting high central molecular gas 
surface density ($\Sigma_{H_2}$), appears to be associated with enhanced
SFE. Unfortunately, morphological information for the pre-merger systems in
the present sample is sparse, with most of them classified as peculiars or 
irregulars. Therefore, without further data it is difficult to explore
trends in the SFE evolution with bulge-to-disk ratio. 
Moreover, the orbital dynamics of the encounter also play a  role,  albeit
a modest one, in regulating the gas inflow and therefore the peak of
star-formation activity  (Mihos \& Hernquist 1996). In particular, prograde
encounters produce gas inflows at early stages, as opposed to retrograde  
ones, where the gas dissipation occurs close to the final stages of the
interaction. 
To further explore this trend, kinematic information for the interacting
systems in the present sample are required.  
Since interacting galaxies may have a  range of bulge-to-disk  ratios and
different orbital dynamics, a scatter is expected in the evolution of their
SFEs.  
Unfortunately, the present sample cannot be used to assess the relative
importance of galaxy properties and interaction  geometry in moderating
the observed activity.  
The fact that the SFE of the present sample peaks close to nuclear
coalescence (despite the scatter) indicates that systems with late gas
inflows (i.e. bulge dominated galaxies in the Mihos \& Hernquist (1996)
scenario) are likely to be over-represented in our sample (Mihos 1999). A
similar result is obtained by Gao \& Solomon (1999) and Gao et al. (1998)
who studied the SFE as a function of nuclear separation for FIR luminous
galaxies (some of which overlap with the present sample). Therefore, it is
probable that  the FIR selection biases the sample towards  systems with
late gas  inflows. Although the present sample also comprises
morphologically selected galaxy pairs, most of the pre-merger galaxies are
FIR-luminous.   

\begin{figure} 
\centerline{\psfig{figure=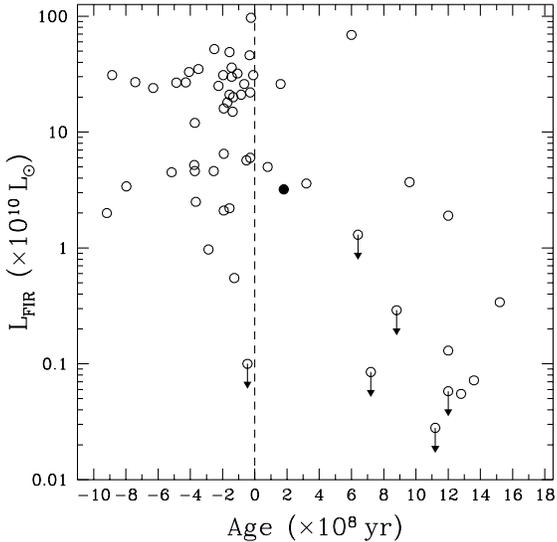,width=0.45\textwidth,angle=0}} 
\caption{ FIR luminosity, $L_{FIR}$, as function of the galaxy
`age' parameter. The points are the same as in Figure \ref{fig_sfe}.
The dashed line separates pre- and post-mergers. For pre-mergers there is
no obvious trend between $L_{FIR}$ and galaxy `age'. However, the $L_{FIR}$ 
at post-merger stages steeply decreases as the the interaction induced
starburst declines. 
}\label{fig_fir}
\end{figure}

\begin{figure} 
\centerline{\psfig{figure=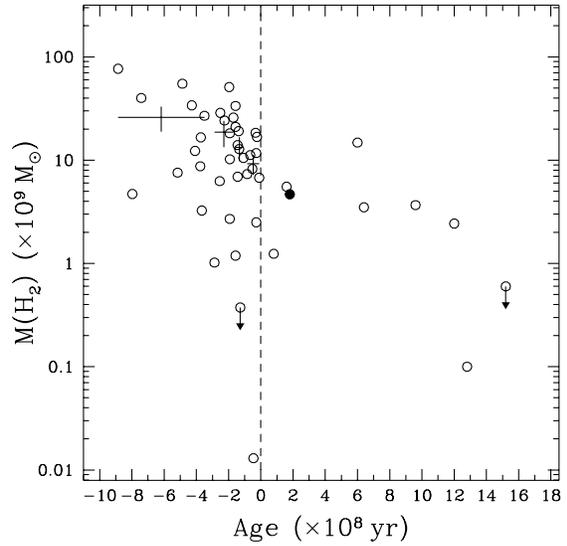,width=0.45\textwidth,angle=0}} 
\caption{Molecular hydrogen mass, $M(H_2)$, as function of the galaxy
`age' parameter. The points are the same as in Figure \ref{fig_sfe}.
The crosses signify the mean $M(H_{2})$ for pre-mergers averaged within
`age'-parameter bins. The horizontal  error bars represent the width of  
the bin, selected so that each bin comprises the same number of
systems ($\approx10$). The dashed line separates pre- and
post-mergers. There is evidence for a decrease in  $M(H_{2})$ along the
merger sequence, likely to be due to $M(H_{2})$ depletion by the
interaction induced starburst.
}\label{fig_mh2}
\end{figure}

Gao \& Solomon (1999) argue that the observed increase in the SFE for
close galaxy pairs is primarily due to the decreasing mass of
available $M(H_2)$ as the interaction progresses to advanced
stages. However, Solomon \& Sage (1988) concluded that the elevated
SFE of strongly interacting/merging galaxies in their sample (compared to  
isolated galaxies) is mainly due to higher FIR luminosities rather than low 
$M(H_{2})$ masses (estimated by their CO luminosities).   
Figures \ref{fig_fir} and \ref{fig_mh2} plot the FIR luminosity and 
molecular hydrogen mass as a function of the `age' parameter respectively
for the  interacting systems in the present sample. 
We find no correlation between `age' and $L_{FIR}$, with a Spearman
rank correlation coefficient $r=0.14$ and a probability for no
correlation $P=36\%$.  On the contrary, an anti-correlation is found
between `age' and $M(H_2)$, albeit a weak one, with $r=-0.32$ and
$P=5\%$, in better agreement with Gao \& Solomon (1999). 
In any case, the  elevated SFE close to nuclear coalescence can be
attributed to star-formation triggered off by the interaction process that
efficiently converts the existing giant molecular clouds into stars (Gao \&
Solomon 1999; Solomon \& Sage 1988). 
However, Solomon \& Sage (1988) also argue that  the observed high SFE in
late interacting systems may arise from an underestimation of the $M(H_2)$
mass in these galaxies by the CO  luminosity. In particular, in high
density environments, similar to those expected in the nuclear regions of
merging galaxies, the CO--to--H$_2$ conversion factor might be
significantly higher than that in  Galactic Giant Molecular Clouds
(GMCs). Moreover, a metallicity lower than that of the Galactic GMCs would
also significantly reduce the sensitivity  of the CO emission to molecular
hydrogen and thus result in underestimation  of the $M(H_2)$ (Combes 1999
and references therein). Nevertheless, dynamical arguments suggest that the
Galactic CO--to--H$_2$ conversion factor is likely to give close to
correct or  even  underestimate  H$_2$ masses, even in the extreme
environments found in  merging galaxies (Solomon \& Sage 1988; Solomon 
et  al. 1997; Downes et al. 1993).   
Additionally, the fact that
Figure \ref{fig_sfe} exhibits less scatter than that in Figures
\ref{fig_fir} and  \ref{fig_mh2} implies that the increase in SFE close to
nuclear  coalescence in Figure \ref{fig_sfe} is likely to be real.  
 
Additionally, it is also clear from Figures \ref{fig_fir}   and
\ref{fig_mh2} that at later, post-merger times, the FIR luminosity
decreases as the star-formation declines, while the molecular hydrogen is
further depleted  but at a slower rate. Consequently, the SFE of merger
remnants declines after the merger event.    

Also shown in Figure \ref{fig_sfe} are the typical SFEs for isolated
spirals (Solomon \& Sage 1988) and  normal ellipticals (Lees et
al. 1991). For ellipticals, the presence of few systems with high SFE in
the Lees et al. (1991) sample biases the mean to large values. A more
robust estimator of the central value of a distribution with a long tail is
the median value. Additionally, the presence of $M(H_{2})$ upper limits in
the Lees et al. (1991) sample requires the use of survival analysis to
estimate statistical quantities. The median SFE for ellipticals is
therefore, estimated using  the {\sc asurv} Rev. 1.2 code (Feigelson,
Isobe \& LaValley 1992), which implements the methods presented in
Feigelson \& Nelson (1985).    

Early, well separated interacting  systems have SFE comparable to that  of
isolated spirals, suggesting that these systems are in a pre-starburst stage
(Lo, Gao \& Gruendl 1997). Additionally, there is also evidence that 
the SFE of merger remnants and `normal' ellipticals form a continuous
decreasing sequence. However, CO(1-0) observations of a larger sample of
merger remnant candidates is needed to further explore their association with
ellipticals. Casoli et al. (1991) and Hibbard \& van Gorkom (1996)
also studied the evolution of the SFE  of a small sample of pre- and
post-merger. They found that post-mergers have SFE that
closely resembles that of `genuine' ellipticals, although they are
relatively rich in cold gas (molecular and neutral hydrogen) compared to 
E/S0s. Nevertheless, Hibbard \& van Gorkom (1996) argue that these merger 
remnants are likely to rid a large fraction of their gas within few Gyr,
mainly due to modest on-going star-formation.

In Figure \ref{fig_sfe} there is evidence that the  ring galaxy ESO
034-IG11 has lower SFE compared to  merger remnant candidates produced by
major disk-galaxy interactions. Horellou et al. (1995) also found a mean  
$SFE=16\pm10\,\mathrm{L_{\odot}/M_{\odot}}$ for ring galaxies, suggesting
that the 
star-formation activity in ring systems declines faster after the close
approach of the intruder galaxy compared to disk-galaxy mergers. 
Alternatively, this might indicate that  the encounters that give rise to
ring systems do not produce as powerful starbursts as disk-galaxy
mergers. Indeed, the star-formation activity in ring systems is mainly
restricted in the ring (Higdon \& Wallin 1997) due to the propagation of
density waves, rather than the nuclear region as in disk-galaxy 
mergers. Therefore, the density waves triggered by head-on collisions are
less likely to produce the high concentrations of molecular hydrogen found
in the nuclear regions of the systems resulting from disk-galaxy 
encounters. As demonstrated in Figure \ref{fig_sh2sfe}, such high
concentrations of molecular hydrogen  are also associated with powerful
starbursts.

\subsection{Molecular hydrogen surface density}

\begin{figure} 
\centerline{\psfig{figure=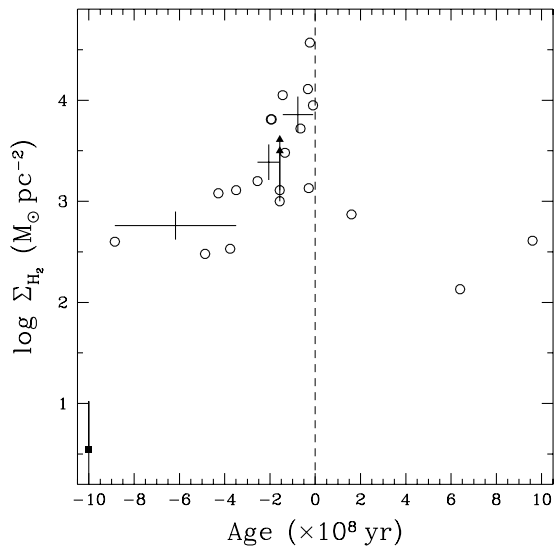,width=0.45\textwidth,angle=0}} 
\caption{ Central nuclear surface density of molecular hydrogen,
$\Sigma_{H_2}$, as a function of the galaxy `age' parameter. The points are
the same as in Figure \ref{fig_sfe}. The filled squares corresponds to the
mean $\Sigma_{H_2}$ of isolated spirals (Kennicutt 1998; however see
discussion in text). The crosses signify the mean $\Sigma_{H_2}$ for
pre-mergers averaged  within  `age'-parameter bins. The horizontal error
bars represent the width  of each bin, selected so that each bin comprises
about 5 points. The vertical  error bars are the standard error around the
mean  $\Sigma_{H_2}$ within each bin.  There is evidence that the
$\Sigma_{H_2}$ for pre-mergers peaks close to nuclear coalescence,
indicating gas inflows arising from gravitational instabilities. At
post-merger stages the  $\Sigma_{H_2}$ seems to decline but the poor
statistics do not allow any firm conclusions to be drawn. 
}\label{fig_sh2}
\end{figure}

The nuclear surface density of molecular hydrogen, $\Sigma_{H_2}$, is
plotted as a function of the galaxy `age' parameter in Figure 
\ref{fig_sh2}. There is evidence for increasing $\Sigma_{H_2}$ as the
interaction progresses towards the final nuclear coalescence. Indeed,
we find a Spearman rank correlation coefficient $r=0.78$,
corresponding to a probability that the observed trend arises by
chance $P=0.2\%$. However,
there is significant scatter and the data are sparse, since there is still
limited number of galaxies with high resolution CO(1-0)
observations. An additional caveat is that the synthesised beam (i.e. the
minimum resolving element) probes regions of different linear size for
systems at different distances, contributing to the observed scatter.   

As explained in the previous section, we attempt to compensate for projection
effects by estimating the mean $\Sigma_{H_2}$ within `age' bins of variable
width. The results are also shown in Figure \ref{fig_sh2},
indicating an increase of the mean $\Sigma_{H_2}$ for pre-mergers towards
nuclear coalescence. A similar result was obtained by Gao et al. (1998) who 
investigated the evolution of $\Sigma_{H_2}$ as a function of nuclear
separation of interacting systems. Also shown in Figure \ref{fig_sh2} is
the mean $\Sigma_{H_2}$ for isolated spirals (Kennicutt et al. 1998). It is
clear that the interacting systems have significantly higher $\Sigma_{H_2}$ 
compared to isolated spirals. 
However, it should be noted that the molecular gas surface densities of
isolated spirals are calculated 
by averaging the CO(1-0) emission over the optical radius of the galactic
disk, rather than the nuclear region. Therefore, the $\Sigma_{H_2}$
of field spirals calculated by Kennicutt et al. (1998) is expected to
significantly underestimate their central surface density. We were also
unable to find a representative average $\Sigma_{H_2}$ for ellipticals in
the literature.

Numerical simulations (e.g. Mihos \&  Hernquist 1996) have demonstrated
that tidal encounters trigger significant gas in-flows that lead to high 
central concentrations of gas. However, as discussed in the previous
section, these models also predict that gas dissipation occurs at different
stages of the interaction depending  on the galaxy internal structure and 
interaction geometry. 

For post-mergers the situation is less clear, since there are only three
merger remnants in our sample with available high resolution CO(1-0)
observations and thus, any conclusions are hampered by poor
statistics. Nevertheless, there is evidence for decreasing  $\Sigma_{H_2}$
as the system evolves after the nuclear coalescence.  
Observations of the molecular gas distribution of a statistically complete
sample of merger remnants are essential to further explore this
trend. Moreover, little is known  about the molecular gas distribution of
`normal' elliptical galaxies. Comparison between the $\Sigma_{H_2}$ for
ellipticals and candidate merger remnants is essential to test this aspect
of the merging  hypothesis.  

\subsection{Cold gas}

\begin{figure} 
\centerline{\psfig{figure=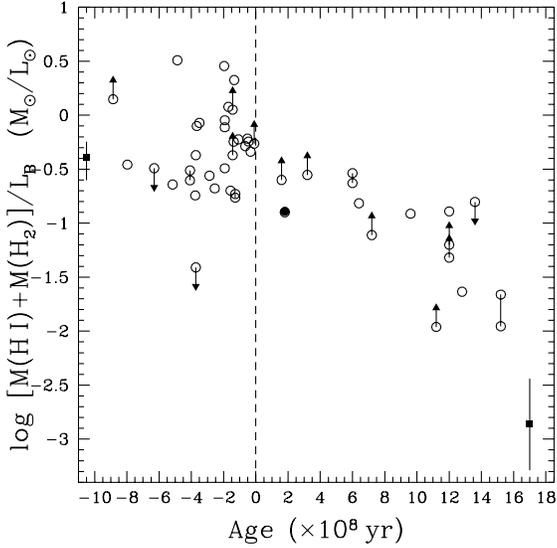,width=0.45\textwidth,angle=0}}
\caption{ Total mass of neutral and molecular hydrogen normalised to the 
$B$-band luminosity as a function of galaxy `age'. The points are the same as
in Figure \ref{fig_sfe}. Points connected with a line represent the upper
and lower $[M(H_2)+M(HI)]/L_{B}$ limits for that system. For isolated
spirals and ellipticals (filled squares) the mean  $[M(H_2)+M(HI)]/L_{B}$
is taken from Young \& Knezek (1989)  and Bregman, Hogg \& Roberts (1992)
respectively. The filled circle is the ring system ESO 034-IG11.
There is evidence for a decrease in  $[M(H_2)+M(HI)]/L_{B}$
from pre- to post-mergers, likely to be due to cold gas depletion. 
}\label{fig_mt}
\end{figure}

The ratio of cold (molecular and neutral hydrogen) gas mass to the blue
band luminosity ($[M(HI)+M(H_2)]/L_{B}$) is plotted as a function of the 
`age' parameter in Figure \ref{fig_mt}. This ratio estimates  the
fraction of cold gas mass in the system. There is evidence 
that $(M(HI)+M(H_2))/L_{B}$ is, on average, decreasing along the merging
sequence from early interacting systems to late merger remnants, indicating
cold gas depletion during the interaction. For post-mergers, we find
that `age' and $(M(HI)+M(H_2))/L_{B}$ are anti-correlated with a
Spearman rank coefficient $r=-0.86$ and a probability that the
distribution is uniform $P=0.3\%$. 

\begin{figure} 
\centerline{\psfig{figure=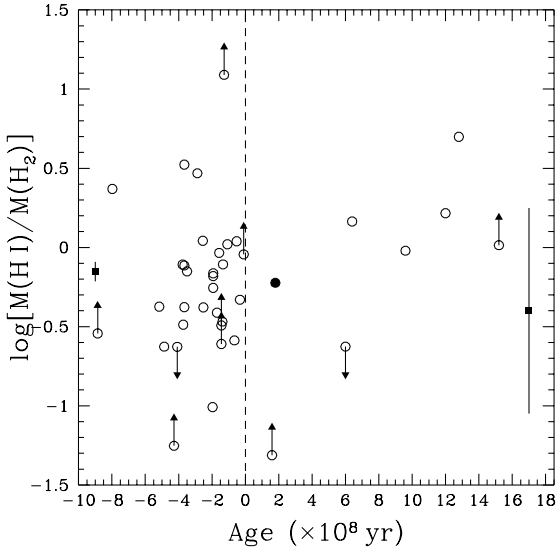,width=0.45\textwidth,angle=0}} 
\caption{Neutral to molecular hydrogen mass, $M(HI)/M(H_2)$, 
as a function of galaxy `age'. The points are the same as in Figure
\ref{fig_sfe}. For isolated spirals and ellipticals (filled squares) the
mean $M(HI)/M(H_2)$ is taken from Young \&   Knezek (1989)  and Wiklind
et al. (1995) respectively. The filled circle is the ring system ESO
034-IG11. There is no obvious trend, implying little net 
conversion from H\,I to H$_{2}$ during the interaction (however see
discussion in text). 
}\label{fig_mhimh2}
\end{figure}

As explained in the previous section, gravitational instabilities during
the interaction drive most of the gas into the centre of the system,
where it is likely to be efficiently converted into stars. 
Additionally, Hibbard \& Van Gorkom (1996) found little evidence for
neutral hydrogen within remnant bodies, with most of it lying in the
outer regions (i.e. tidal features).  
Numerical N-body simulations have shown that the gravitational forces
experienced during the merger can force about half of the outer disc
H\,I  into a tail, the rest of the H\,I being forced into 
the inner regions (Hibbard \& Mihos 1995). As less than a quarter of the
total H\,I is found within these regions, Hibbard \& Van Gorkom (1996)
concluded that most of the centrally forced H\,I gas is converted during the
merger into some other form. They propose that the gas has been turned into
molecular gas, stars or has been  heated up to X-ray temperatures,
either through compression leading to cloud-cloud  collisions or through
energy input from massive stars and supernovae. 
The presence of Balmer absorption lines in the merger remnants NGC 7252 and
NGC 3921 (Dressler \& Gunn 1983; Schweizer 1996) is direct evidence that
some of the original atomic gas is ultimately converted into stars. 
Searches for molecular hydrogen in these same galaxies have revealed that
although they are gas rich compared to  ellipticals or S0s, they have
below average molecular gas content for their spirals progenitors (Solomon
\& Sage 1988; Young \&  Knezek 1989; Hibbard \& van Gorkom 1996),
indicating that any net conversion of atomic to molecular hydrogen is 
relatively inefficient.
This is also demonstrated in Figure \ref{fig_mhimh2}, where we plot the
ratio of neutral to molecular hydrogen mass as a function of the `age'
parameter. It is clear that there is significant scatter without any
obvious correlation, suggesting that the net conversion  from H\,I to H$_2$
is not large during the merging. However, it should be noted that any 
conversion from H\,I to H$_{2}$ is likely to take place in the central
galaxy regions, whereas Figure \ref{fig_mhimh2} plots the global gas
properties of interacting systems. 

Regarding the X-ray properties of mergers,  Read \& Ponman (1998) found  an
increase in the X-ray luminosity of galaxies close to the nuclear
coalescence indicative of the presence of hot gas. However, they found 
little evidence for the presence  of hot X-ray emitting gas in merger
remnants.  They concluded that this is likely to be due to galactic winds,
similar to those observed in the nearby starburst M\,82, that blow
the hot gas out of the system.

Elliptical galaxies are also known to have little cold gas. This is
demonstrated in  Figure \ref{fig_mt} showing the mean
$[M(HI)+M(H_2)]/L_{B}$  ratio for ellipticals (Bregman, Hogg \& Roberts 
1992).  Although merger remnants are gas rich compared to normal
ellipticals, they seem to form a sequence that supports the
merger scenario as a possible formation mechanism for elliptical galaxies. 
Hibbard et al. (1994) studied the gas properties of the merger remnant NGC
7252 and found that although it is gas rich compared to ellipticals and
S0s, it is likely to resemble these galaxies in few Gyrs. In particular,
the H\,I is mostly found in the tidal tails, while the atomic gas content
of the remnant body is typical to that of E/S0s. There is also some
evidence for on-going conversion of the returning tidal tail H\,I into
stars. Moreover, the molecular gas in the NGC 7252 is also likely to be
depleted within the next few Gyrs, due to modest on-going star-formation. 

The ring galaxy ESO 034-IG04 in Figure \ref{fig_sfe} has  cold gas mass
fraction similar to that of merger remnant candidates resulting from major
disk-galaxy encounters. A similar result was obtained by Horellou et
al. (1995), who found a mean $[M(HI)+M(H_2)]/L_{B}$  ratio of $0.22\pm0.17$
for ring systems.

\subsection{Radio flux density}

\begin{figure} 
\centerline{\psfig{figure=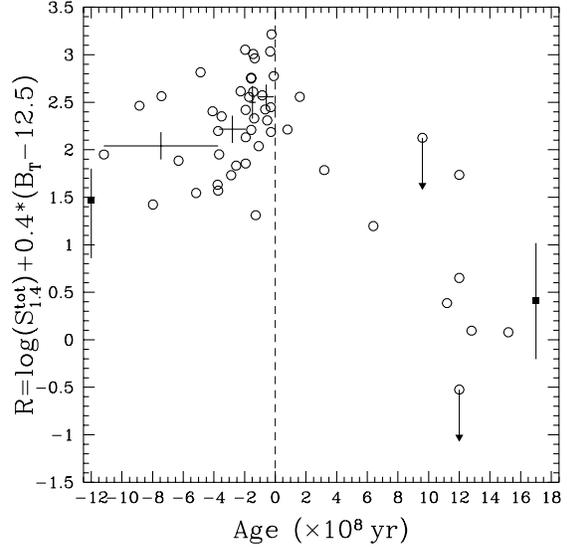,width=0.45\textwidth,angle=0}} 
\caption{Total radio to $B$-band flux ratio
$(R=\log S^{tot}_{1.4}+0.4\times[B_{T}-12.5])$ as a function of galaxy   
`age'. The points are the same as in Figure \ref{fig_sfe}. The filled
squares are the mean  $R$ for isolated spirals (Klein 1982) and ellipticals
(Sadler 1984).
The crosses signify the mean $R$ for pre-mergers averaged within
`age'-parameter bins. The horizontal error bars represent the width of the
bin, selected so that each bin comprises about 10 systems. The vertical
error bars are the standard error on the mean.  Although the mean $R$ for
pre-mergers is higher that that of isolated spirals, it marginally
increases along the merger sequence towards the nuclear coalescence. The
$R$ parameter at post-merger stages declines steeply, to values typical of
normal ellipticals. 
}\label{fig_r}
\end{figure}

We define the ratio, $R$, between total radio (1.4\,GHz) flux density,
$S^{total}_{1.4}$ and $B$-band luminosity (Hummel 1981)

\begin{equation}\label{eq_r}
R=\log(S^{total}_{1.4})+0.4\times(B_{T}-12.5).
\end{equation}

\noindent The $R$ parameter is independent of distance and estimates the
ratio between radio power and optical luminosity. It has
been demonstrated that the mean radio power is proportional to the  mean
optical luminosity of galaxies (Hummel 1981). The $R$ parameter
also takes into account this effect, providing an estimate of the excess
radio emission in galaxies due to star-formation or AGN activity.  

The $R$ parameter is plotted against the galaxy  `age' in Figure
\ref{fig_r}. For pre-mergers, we find a marginally significant
correlation with Spearman rank coefficient
$r=0.41$ and a probability that the distribution is uniform
$P=1.3\%$. The mean $R$
within different `age' bins (of variable width) 
is also shown in the same figure.  In agreement with our previous
result the mean $R$ parameter for 
pre-mergers, although on average higher than that of isolated spirals
($\approx0.8$\,dex; Klein 1982), marginally increases along the merging
sequence towards the final stages of the tidal encounter.  This is a
surprising result since the radio  flux is related to star-formation
activity in galaxies (Condon et al. 1992  and references therein). This can
be partly  attributed to projection  effects and  differences in the
details of individual interactions. Moreover, it has been shown that tidal
encounters primarily act to increase the nuclear galaxy star-formation
within the central kpc (Keel et al. 1985; Kennicutt et al. 1987) and only 
moderately affect the activity (i.e. star-formation) in the disk. In
particular, Hummel (1981) found little difference between the disk radio
power (normalised  to the blue-band luminosity) of interacting pairs and
isolated spirals. On the contrary an increase by a factor of 2.5 was found
for the central radio power of the  two samples.  Also shown in Figure
\ref{fig_r} is the mean $R$ for ellipticals (Sadler 1984). It is clear that
the radio properties of merger remnant candidates and ellipticals are in
fair agreement. 
 
\begin{figure} 
\centerline{\psfig{figure=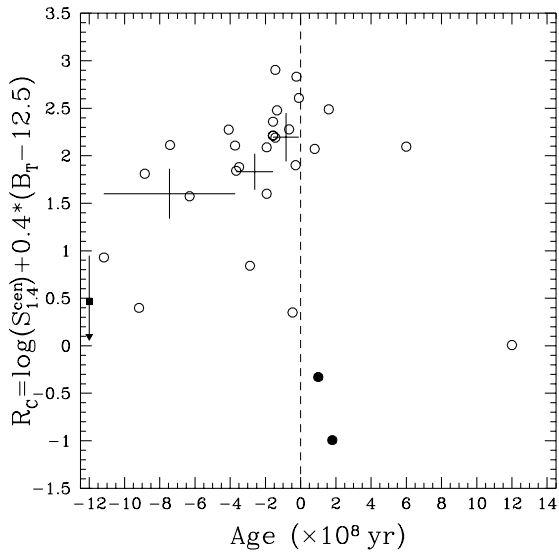,width=0.45\textwidth,angle=0}} 
\caption{Central radio to the $B$-band flux ratio
$(R_{C}=\log S^{cen}_{1.4}+0.4\times[B_{T}-12.5])$ as a function of
galaxy  `age'. The points are the same as in Figure \ref{fig_sfe}. The
filled square is the mean  $R_{C}$ for isolated spirals (Hummel 1981). 
The filled circles are the ring galaxies ESO 034-IG04 and ESO 138-IG29.
The crosses signify the mean $R$ for pre-mergers averaged within
`age'-parameter bins. The horizontal error bars represent the width of the
bin, selected so that each bin comprises about 5 points. The standard
error around the mean is shown by the vertical
error bars.  Although the mean $R_C$ for pre-mergers is significantly
higher that that of isolated spirals, it marginally increases along the
merger sequence towards the nuclear coalescence. At post-merger stages
there is evidence for a decline in the $R_{C}$ but the poor
statistics hamper any interpretation.
}\label{fig_rh}
\end{figure}

To further explore changes in the central radio activity along the merging
sequence, we estimate the 1.4\,GHz radio flux density of the galaxies in
the present sample within the central $\approx2$\,kpc diameter region. For
that purpose we employ high resolution radio data available on NED, mostly
taken from Condon et al. (1991). For galaxies without available 1.4\,GHz data,
$S^{cen}_{1.4}$ is estimated from high resolution
observations at other frequencies  (8.4\,GHz or 4.9\,GHz), if available,
assuming a power-law  spectral energy distribution
$S_{\nu}\propto\nu^{-\alpha}$. The radio spectral index is either taken to
be 0.8 or calculated  from the 8.4\,GHz and  4.9\,GHz radio flux densities,
if available. The results are presented in Figure \ref{fig_rh}, plotting
the central radio flux density to blue band luminosity, $R_C$, against   
galaxy `age'. For pre-mergers, we find a Spearman rank coefficient
$r=0.55$ and a probability that the distribution is uniform
$P=0.9\%$, suggesting that  `age' and $R_C$ are marginally
correlated. Similarly, the increase in the mean $R_C$ along the merger 
sequence for pre-mergers is marginal. However, the poor
statistics do not allow any firm  conclusions to be 
drawn. Nevertheless, the mean $R_{C}$ of interacting systems is 
significantly elevated compared to that of isolated spirals
($\approx1.5$\,dex), in agreement with previous studies (Hummel 1981).

The ring galaxies ESO 034-IG11 and ESO 138-IG29 in Figure \ref{fig_rh} have
central radio to the $B$-band flux ratios significantly lower than merger
remnants resulting from major disk-galaxy encounters. This is likely to be
due to the fact that the star-formation in ring systems is concentrated in
the ring rather than the nuclear region, due to density waves triggered by
the nearly head-on collision. 

\section{The FIR--radio correlation}

The logarithmic FIR--to--radio flux density ratio is defined by (Condon et 
al. 1991) 

\begin{equation}
q=\log[(S_{FIR}/3.75\times10^{12})/S^{tot}_{1.4}],
\end{equation}

\noindent where $S_{FIR}$ is the FIR flux density (section \ref{sample},
equation \ref{eq_fir}) and $S^{tot}_{1.4}$ is the total radio flux density
(section \ref{sample}) in units of $\mathrm{W\,m^{-2}\,Hz^{-1}}$. Radio and 
FIR selected starbursts as well as optically selected spiral and
irregular galaxies exhibit a very narrow $q$-distribution 
($\sigma_q\approx0.2$) centred on $<q>\approx2.34$. This tight
distribution is attributed to star-formation activity resulting
in both FIR emission and supernovae explosions, whose remnants emit at
radio wavelengths via sychnotron radiation.  Additionally, galaxies with
radio emission powered by an AGN have, on average,  $q<2$, implying that
the FIR--radio flux ratio can be employed, to the first approximation, 
to constrain the nature of the energising  source (AGN/star-formation;
Condon 1992 and references therein). However, discriminating between
AGN and star-formation activity as the dominant energising source is
not an easy task, especially in the case of dusty FIR-luminous
systems. Nevertheless, recent studies suggest that mid-infrared
spectroscopy can provide an efficient tool for classifying the energy
source that dominates the observed activity (Genzel et al. 1998; Lutz 
et al. 1999; Rigopoulou et al. 1999). 

\begin{figure} 
\centerline{\psfig{figure=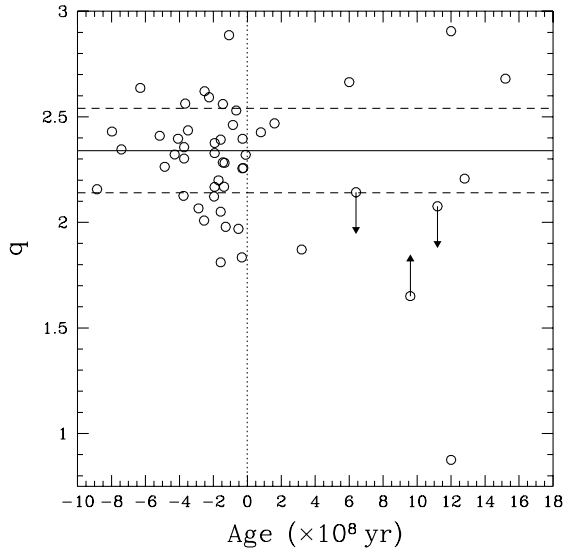,width=0.45\textwidth,angle=0}} 
\caption{Logarithmic FIR--radio flux ratio, $q$, as a function of the
galaxy age parameter. Open circles are the interacting galaxies studied
here.  The continuous line is the mean $q$ for starbursts, while the dashed
lines  signify the $1\sigma$ envelope around the mean. The dotted line
separates pre- from post-mergers. Most of the interacting galaxies are
consistent with star-formation being their energising source. 
}\label{fig_q}
\end{figure}

Figure \ref{fig_q} plots $q$ as a function of the galaxy age
parameter. Also shown in this figure is the region occupied by starbursts
and normal spirals. The interacting systems  in the present sample (both
pre- and post-mergers) have a mean FIR--radio flux ratio $<q>=2.36\pm0.25$,
in fair agreement with the canonical value of $q=2.34\pm0.20$. Therefore,
the $q$ parameters of most of these systems are consistent with
star-formation activity being the main energising source. 
There also few pre-mergers in the sample with $q<2$, which might be
indicating the presence of an AGN contributing to the observed
activity. Interestingly, all these systems have nuclear 
separation  $\delta r<5$\,kpc implying that they are likely to be at an
advanced interaction stage, close to the final merger event. Nevertheless,
the majority of the very close galaxy pairs ($\delta r<5$\,kpc) have
FIR--radio flux ratios consistent with star-formation  activity.    

Smith et al. (1993) also investigated the FIR--radio correlation of
interacting galaxies and concluded that star-formation is likely to be the
main source responsible for the observed FIR and radio activity.
Similarly,  Bushouse et al. (1988) studied the FIR properties of interacting
galaxies and concluded that it is not necessary to invoke mechanisms other 
than starbursts to account for their activity.  Dahari (1985) used optical
spectra to determine the nature of the 
energising source in interacting systems and found little evidence for an
excess of Seyfert nuclei in paired galaxies compared to isolated
spirals. Moreover, he found no Seyfert type spectra in a sub-sample 
of extremely distorted spirals (e.g. similar the majority of pre-mergers
studied here). 
Genzel et al. (1998) and Rigopoulou et al. (1999) used mid-infrared
spectroscopy to explore the nature of ultra-luminous infrared galaxies
(ULIRGS), many of which are experiencing on-going interactions. They
conclude that starburst activity  dominates the bolometric luminosity
of the majority of these extreme systems, although many of them also
have an AGN component. Moreover, Rigopoulou et al. (1999) found no
evidence for an increase in AGN-dominated ULIRGS with decreasing  
projected nuclear separation.

Merger remnant candidates in Figure \ref{fig_q} exhibit significant scatter,
with many of them deviating from the expected relation for
starbursts, although the poor statistics do not allow any firm conclusions
to be drawn. 
Moreover, little is known about the FIR-radio flux ratio distribution of
merger remnants. Nevertheless, elliptical galaxies, with which post-mergers
are most likely associated, follow the FIR--radio relation
for star-forming galaxies (Wrobel \& Heeschen 1991). Nevertheless, a number
of ellipticals are also found to deviate from this relation having low $q$
values, indicating the presence of an AGN. Moreover, Wrobel \& Heeschen
(1991) also found that a fraction of the elliptical galaxies in their
sample  lied well above the canonical FIR-radio relation (high $q$
values). They argue that these systems are likely to have extended
low-surface brightness radio emission, associated with star-formation, that
might remain undetected by the existing observations.     

\section{Conclusions}\label{conclusions}
In this paper we compile a sample of interacting/merging galaxies
aiming to study the evolution of the gas properties and star-formation
along a galaxy merger sequence. The present sample, although not
complete,  is representative of interacting systems and merger
remnants spanning a range of properties. Our conclusions are
summarised below   

\begin{enumerate}

\item We find a statistically significant increase in the SFE of
on-going mergers close to the 
final stages of nuclear coalescence. Nevertheless, there is significant
scatter, attributed to both projection effects and differences in the
interaction  details of individual systems. The observed trend is likely
to be due to  $M(H_2)$ depletion by star-formation. At post-merger stages,
despite the poor statistics, there is evidence that the SFE declines to
values typical to ellipticals, in agreement with the merger hypothesis.   

\item There is also strong evidence for increasing central surface
density of molecular hydrogen close to nuclear coalescence, indicating
gas dissipation due to gravitational instabilities. However,
projection effects and differences in the interaction details of
individual systems contribute to  the observed scatter. 

\item There is evidence for a decrease in cold gas mass fraction (neutral
and molecular hydrogen) along the merging sequence. This attributed to
H\,I conversion into other forms within the body of the system during the
interaction and $M(H_2)$ depletion due to residual star-formation
activity. This trend also seems to support the  merging scenario for the
formation of ellipticals.     

\item The total radio power normalised to the blue-band luminosity,
although higher than that of isolated spirals, marginally increases
along the merger sequence. This is attributed to the fact that interactions
mainly affect the nuclear galaxy activity, whereas 
there is moderate enhancement in the disk star-formation rate. However, a 
similar result is  obtained for the central radio power of interacting
systems (normalised to the blue-band luminosity). Nevertheless, the nuclear
radio to  blue-band luminosity ratio of interacting systems is
significantly  elevated compared to isolated spirals.

\item The FIR--radio flux ratio distribution of interacting galaxies
is consistent with star-formation being the main energising
source. However, there is evidence that some systems might have an AGN  
contributing to the observed activity. 

\end{enumerate}

\section{Acknowledgements}
We thank Chris Mihos and the anonymous referee for valuable comments
and suggestions. This research has made use of the {\sc nasa/ipac}
Extragalactic Database ({\sc ned}), which is operated by the Jet Propulsion
Laboratory, Caltech, under contract with the National Aeronautics and
Space Administration. The Australia Telescope is funded by the
Commonwealth of Australia for operation as a National Facility managed
by {\sc csiro}. The Digitised Sky Survey was produced at the Space Telescope
Science Institute under US government grant NAG W-2166.

\appendix

\section{Notes on individual galaxies with new radio data}

\begin{figure} 
\centerline{\psfig{figure=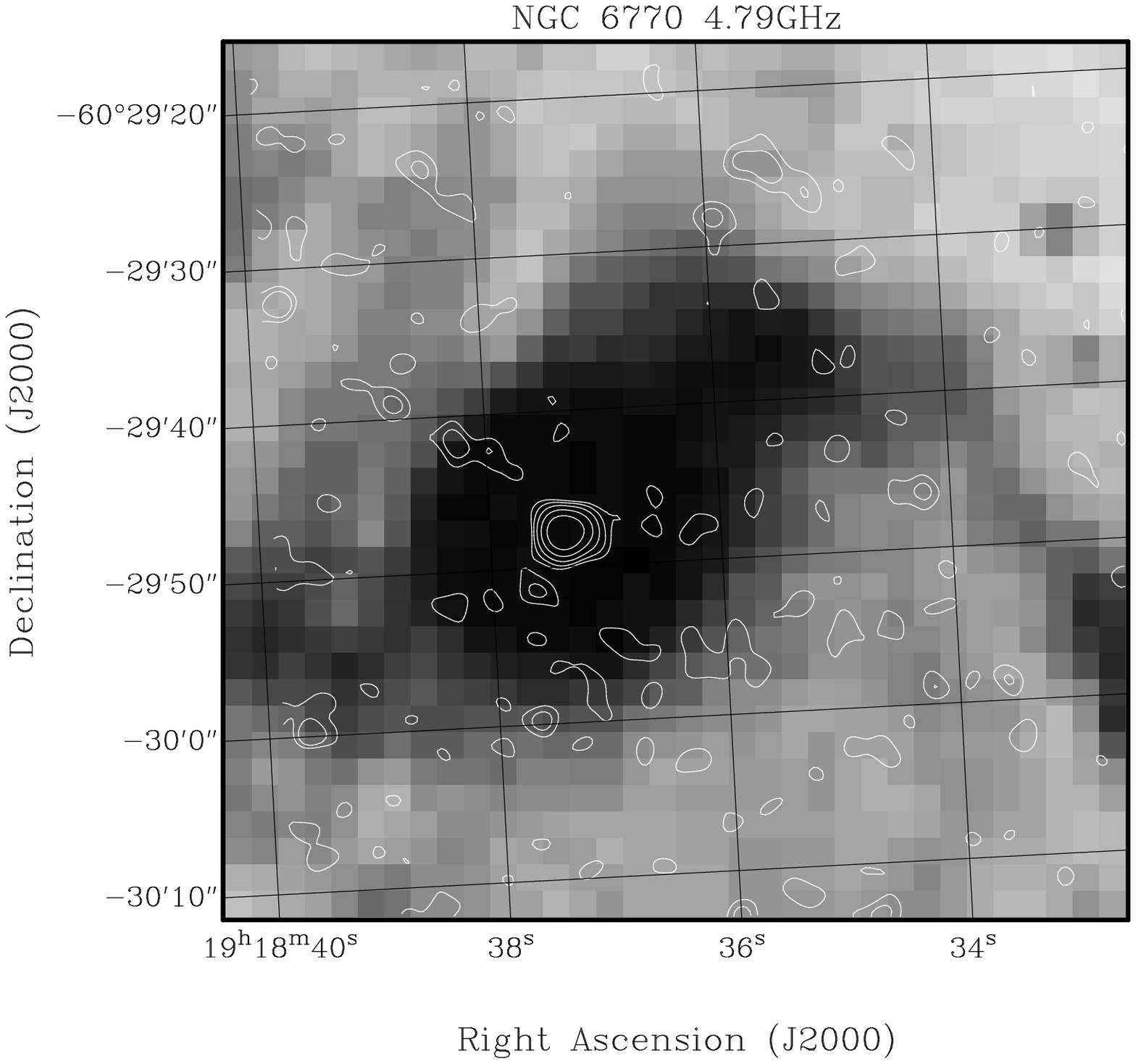,width=0.45\textwidth,angle=0}} 
\caption{NGC 6770. Radio contours at 4.79\,GHz (6\,cm) overlayed on the
Digital Sky Survey optical image (linear scale). The radio contours are
logarithmically spaced between 0.05\,mJy/beam and 1.35\,mJy/beam using a
logarithmic step of 0.48.} 
\label{fig_n6770}
\end{figure}

\noindent{\bf NGC 6769/70} \\ 
This galaxy is strongly  interacting with an equal mass spiral NGC 6770
with a bridge of stars connecting them. The  
optical  centres of the two galaxies lie at  RA = 19$^h$18$^m$22.6$^s$,
Dec.= --60$^{\circ}$30$^{\prime}$03$^{\prime\prime}$ and RA = 19$^h$18$^m$37.6$^s$, Dec.=
--60$^{\circ}$29$^{\prime}$50$^{\prime\prime}$ (J2000) respectively. Surface photometry of the
galaxies is discussed in Storchi \& Patroriza  (1986). The radio contours
at 4.79\,GHz overlayed on the optical image of NGC 6770 are shown in Figure
\ref{fig_n6770}.
\\

\begin{figure} 
\centerline{\psfig{figure=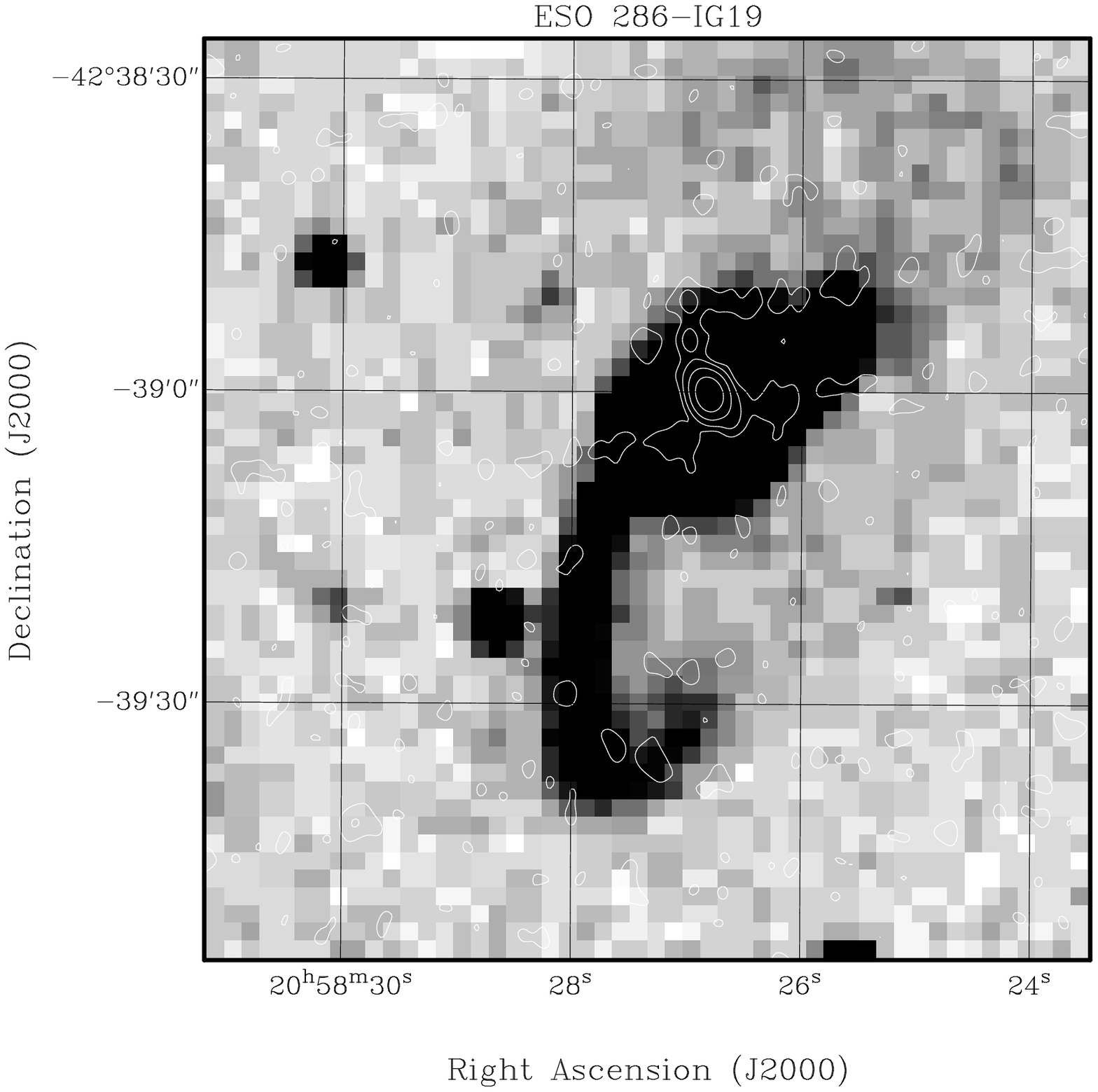,width=0.45\textwidth,angle=0}} 
\caption{ESO 286-IG19. Same as in Figure \ref{fig_n6770}. The radio
(4.79\,GHz) contours are logarithmically spaced between 0.06\,mJy/beam and
3.84\,mJy/beam using a logarithmic step of 0.60.} 
\label{fig_eso286}
\end{figure}

\noindent{\bf ESO 286--IG19} \\ 
Imaging of ESO286--IG19 by Johansson (1991) reveals two tidal tails and a
single r$^{1/4}$ like nucleus. This would tend to  suggest a late stage
merger. Comparison with the sequence of Keel \& Wu (1995) suggests a
dynamical stage of 1.5 (i.e $\approx6\times10^{8}$\,yrs).  The radio
contours at 4.79\,GHz overlayed on the optical image of ESO 286--IG19  are
shown in Figure \ref{fig_eso286}.  
\\

\begin{figure} 
\centerline{\psfig{figure=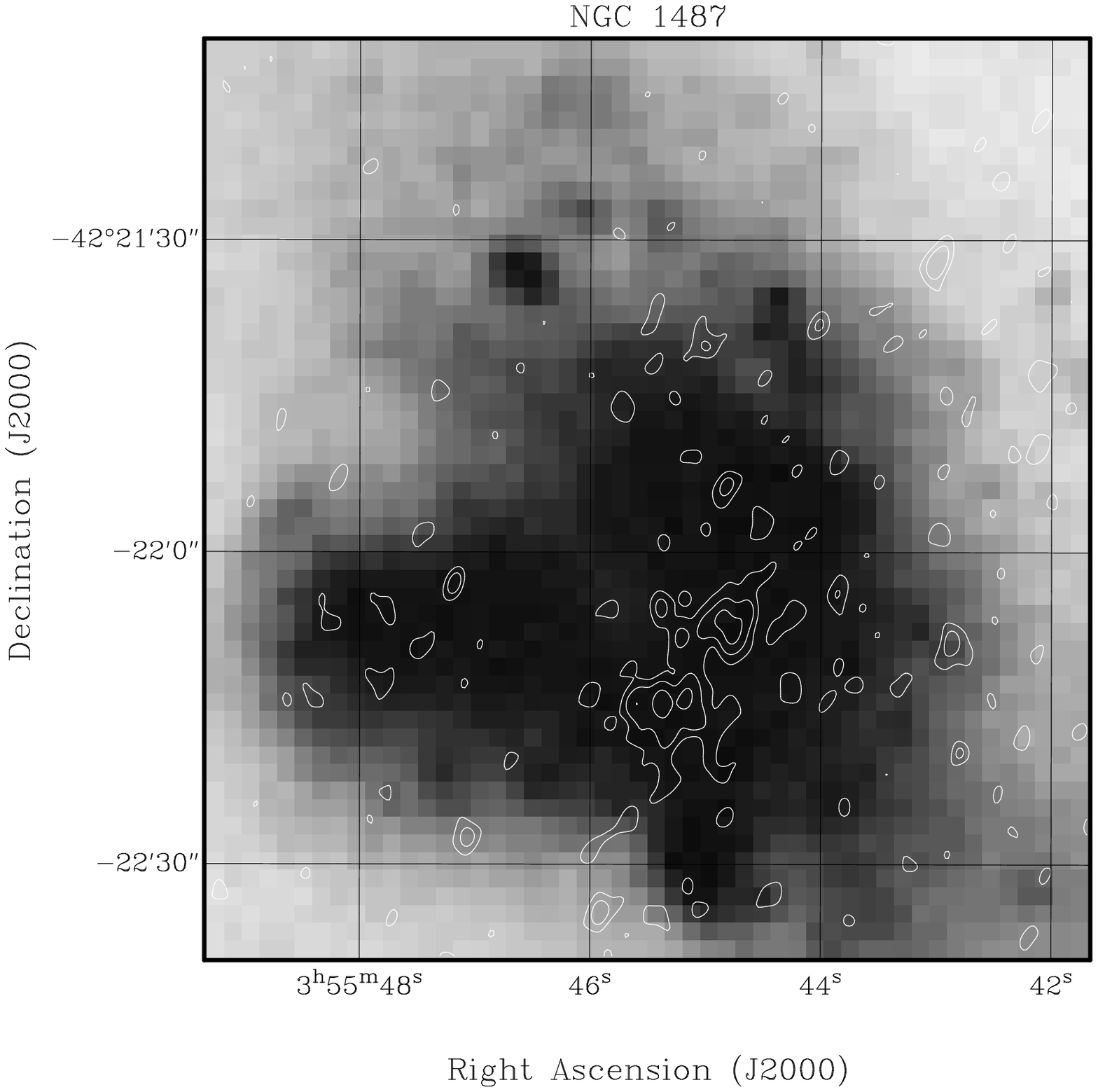,width=0.45\textwidth,angle=0}} 
\caption{NGC 1487. Same as in Figure \ref{fig_n6770}. The radio (4.79\,GHz)
contours are logarithmically spaced between 0.05\,mJy/beam and
0.20\,mJy/beam using a logarithmic step of 0.30.} 
\label{fig_n1487}
\end{figure}

\noindent{\bf NGC 1487} \\ 
This system is in an early stage merger revealing two clear tails and
two nuclei but sufficiently advanced to be one galactic body (e.g. Bergvall
\& Johansson 1995).  It is slightly more evolved than NGC 4676 (``The
Mice'') but less so than NGC 4038/9 (``The Antennae'').  The radio
contours at 4.79\,GHz overlayed on the optical image of NGC 1487  are
shown in Figure \ref{fig_n1487}.  
\\ 

\begin{figure} 
\centerline{\psfig{figure=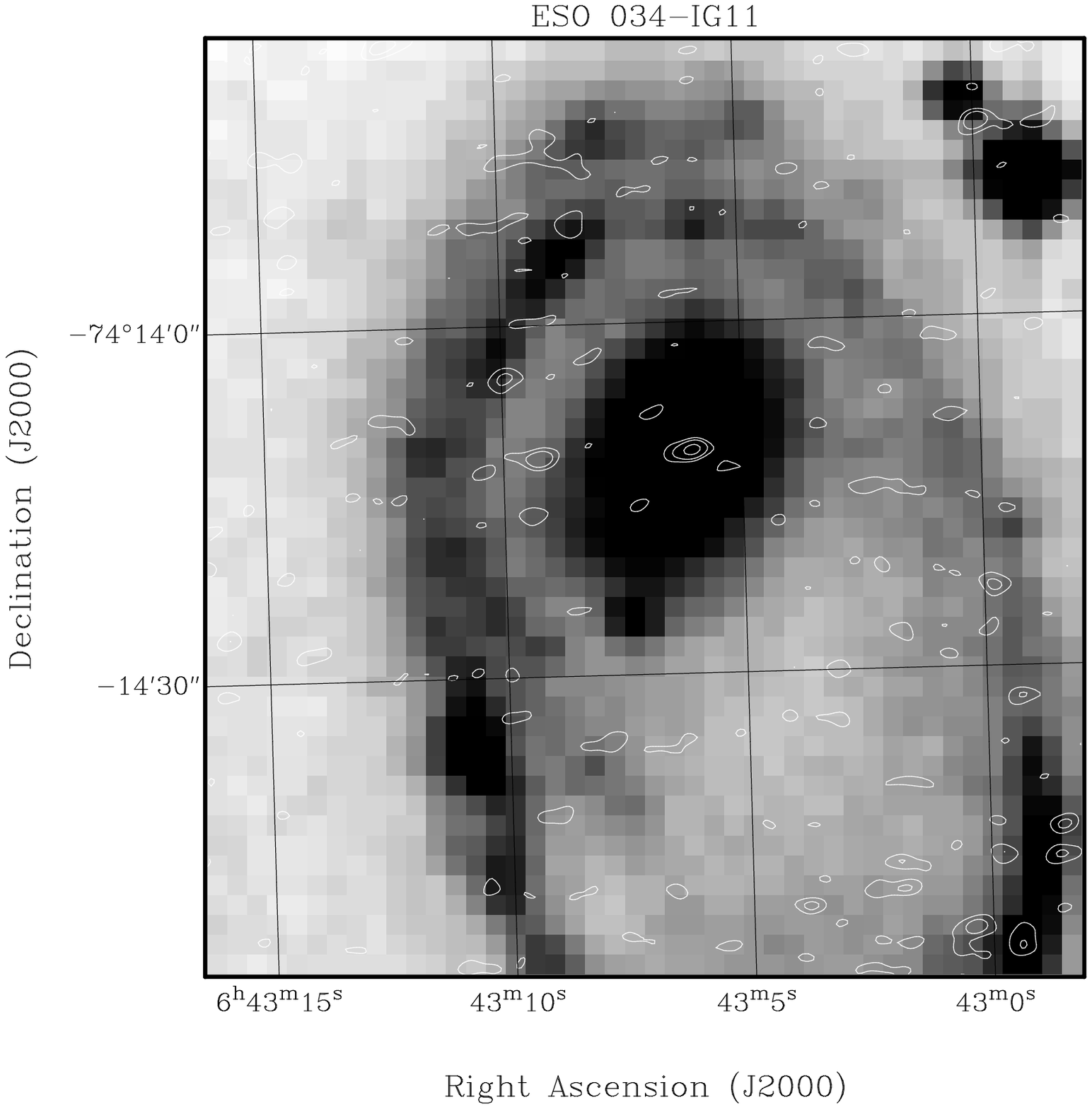,width=0.45\textwidth,angle=0}} 
\caption{ESO 034-IG11. Same as in Figure \ref{fig_n6770}. The radio (4.79\,GHz)
contours  are logarithmically spaced between 0.05\,mJy/beam and
0.20\,mJy/beam using a logarithmic step of 0.30.}
\label{fig_eso034}
\end{figure}

\noindent{\bf ESO 034--IG11} \\ 
Also known as the Lindsay--Shapley Ring (Lindsay \& Shapley 1960). The
asymmetric ring suggests an off--centre collision. It has been studied by
Higdon \& Wallin  (1997). They found an optical bridge from the ring to the
`intruder' galaxy at RA = 06$^h$43$^m$26$^s$, Dec. =
--74$^{\circ}$15$^{\prime}$29$^{\prime\prime}$ (J2000).  The optical nucleus of the perturbed
galaxy  is at RA = 06$^h$43$^m$06.7$^s$, Dec.= --74$^{\circ}$14$^{\prime}$16$^{\prime\prime}$
(J2000). From the expansion rate of the ring in ESO34--IG11 the interaction
occurred about $1.8\times10^8$\,yrs ago
($H_{0}=75\,\mathrm{km\,s^{-1}\,Mpc^{-1}}$; Higdon \& Wallin 1997).   The radio
contours at 4.79\,GHz overlayed on the optical image of ESO 034--IG11  are
shown in Figure \ref{fig_eso034}.   
\\

\begin{figure} 
\centerline{\psfig{figure=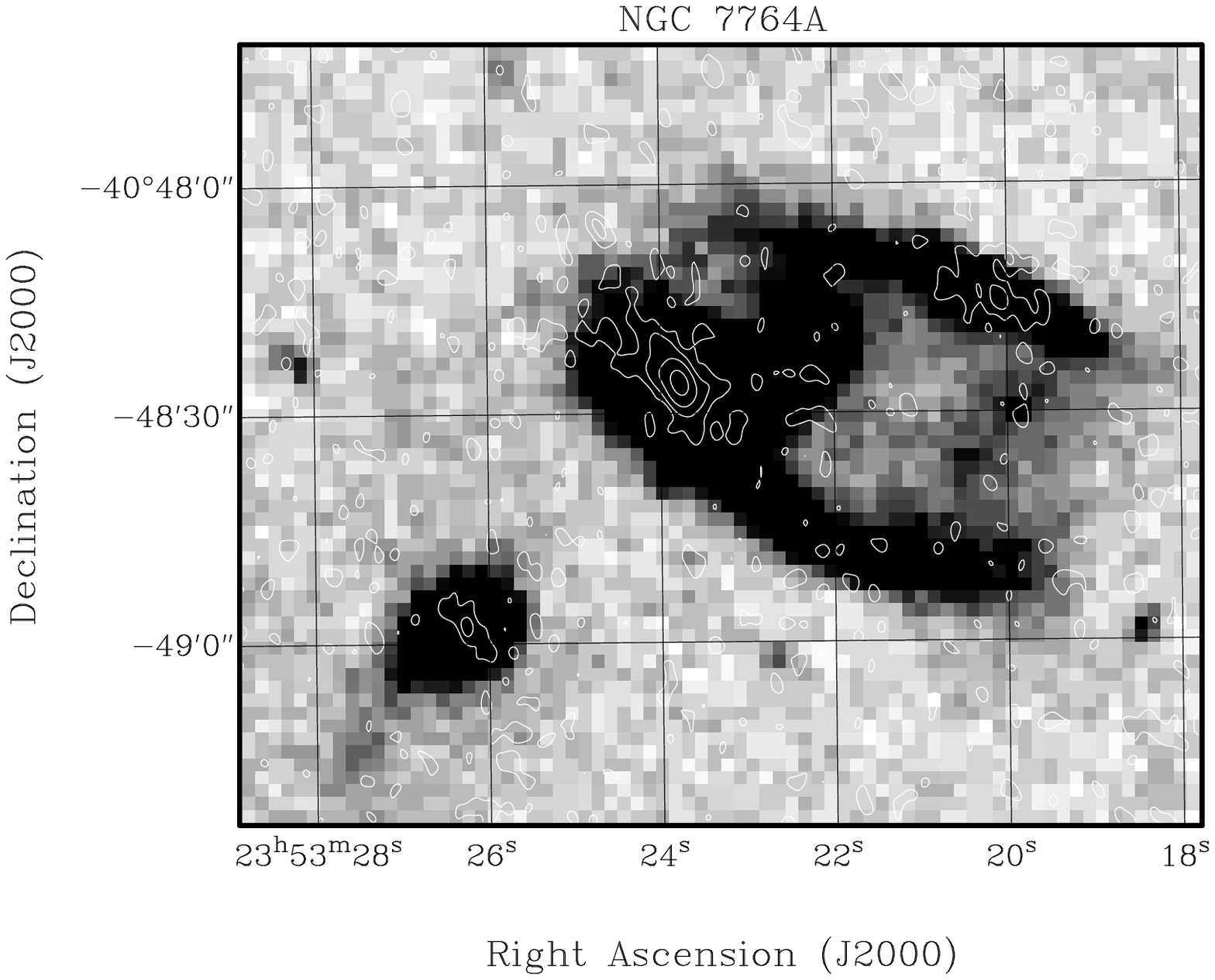,width=0.45\textwidth,angle=0}} 
\caption{NGC 7764A. Same as in Figure \ref{fig_n6770}. The radio
(4.79\,GHz) contours are logarithmically spaced between 0.07\,mJy/beam and
1.89\,mJy/beam using a logarithmic step of 0.48.}
\label{fig_n7764}
\end{figure}

\noindent{\bf NGC 7764A}\\
The NGC 7764A system is an interacting triple system containing NGC 7764A NED2
(AM2350--410), 7764A NED3 to the south--east  and 7764A NED1 to the
north--west. The largest galaxy, NGC 7764A NED2,  
appears to be in the process of merging with NGC 7764A NED1
Borchkhadze et al. (1977) note the
tidal material between these two galaxies. The smallest galaxy, NGC 7764A
NED3 
on the other hand  shows only faint asymmetries in  its outer optical
isophotes.   In this study, the NGC 7764A NED1/NED2 system is treated as a
pre--merger, with two distinct galaxies present.  The radio 
contours at 4.79\,GHz overlayed on the optical image of NGC 7764A  are
shown in Figure \ref{fig_n7764}.  
\\

\begin{figure} 
\centerline{\psfig{figure=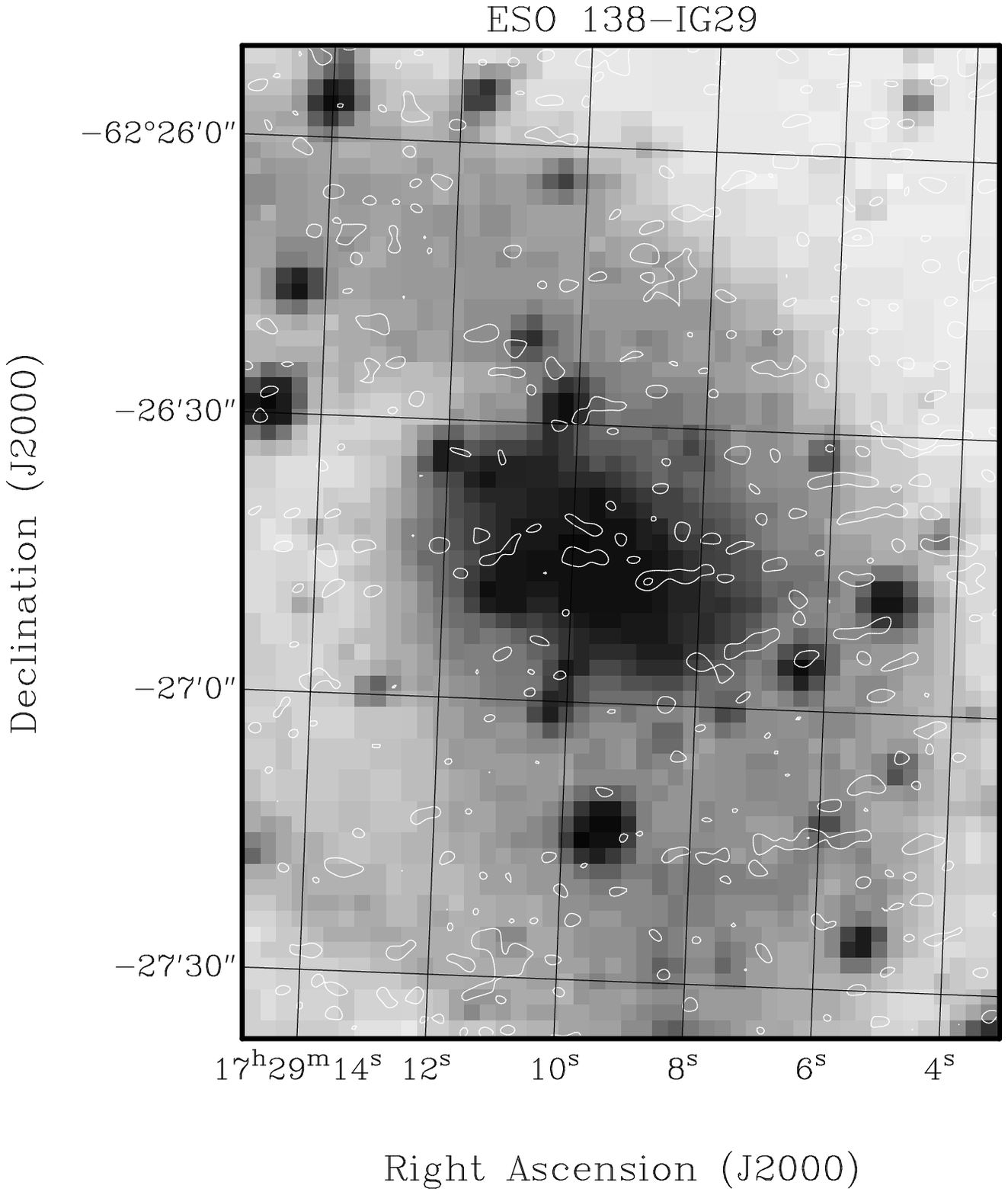,width=0.45\textwidth,angle=0}} 
\caption{ESO 138-IG29. Same as in Figure \ref{fig_n6770}. The radio
(4.79\,GHz) contours are logarithmically spaced between 0.07\,mJy/beam and
0.21\,mJy/beam using a logarithmic step of 0.60.}
\label{fig_eso138}
\end{figure}

\noindent{\bf ESO 138--IG29} \\ 
Also known as the ``Sacred Mushroom'', this galaxy appears to be a young
ring system formed by ESO138--IG30 as it passed through the disk of
ESO138--IG29 less than 10$^8$ yrs ago (Wallin \& Struck--Marcell
1994). Optical imaging and dynamical models of this system have been
carried out Wallin \& Struck--Marcell (1994) which suggest that
ESO138--IG29 was originally an S0 galaxy. The radio
contours at 4.79\,GHz overlayed on the optical image of ESO 138--IG29  are
shown in Figure \ref{fig_eso138}.  
\\ 

\begin{figure} 
\centerline{\psfig{figure=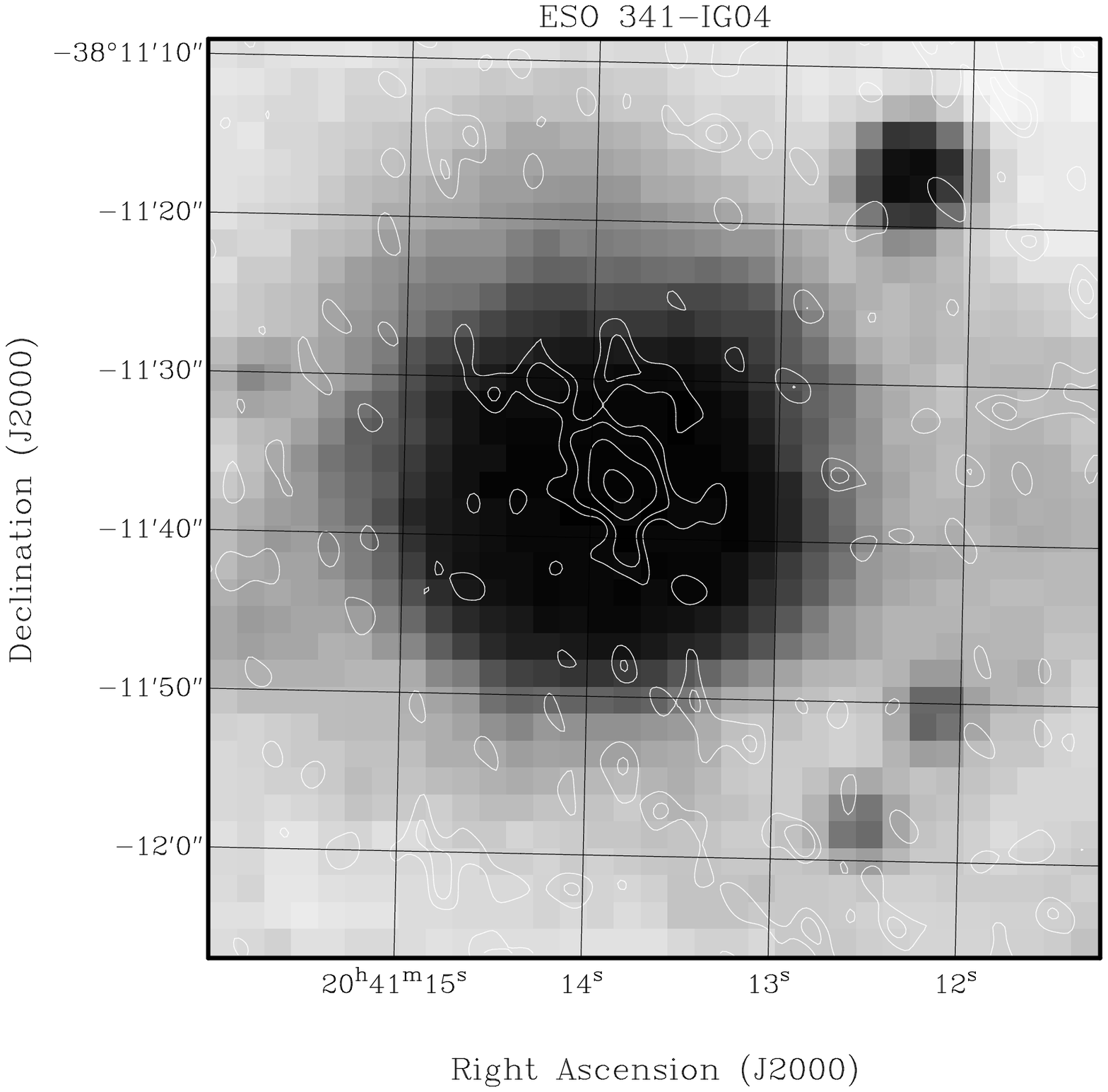,width=0.45\textwidth,angle=0}} 
\caption{ESO 341-IG04. Same as in Figure \ref{fig_n6770}. The radio
(4.79\,GHz) contours are logarithmically spaced between 0.05\,mJy/beam and
0.40\,mJy/beam using a logarithmic step of 0.30.}
\label{fig_eso341}
\end{figure}

\noindent{\bf ESO 341--IG04} \\ 
This galaxy is probably at the  very last stages of a merger. Its optical
appearance is close to that of an elliptical galaxy (its surface brightness
profile follows an r$^{1/4}$ law out to 5 effective radii), although it
still contains a large amount of HI gas (Bergvall et al. 1989). It has one
nucleus and one tail or plume, in addition to the prominent south--west
loop. It appears to be more evolved  than NGC 7252 but not quite at the end
of the Keel \& Wu (1995) merger sequence.  The optical nucleus is at RA
=20$^h$41$^m$14.3$^s$, Dec. =
--38$^{\circ}$11$^{\prime}$40$^{\prime\prime}$ (J2000).   The radio
contours at 4.79\,GHz overlayed on the optical image of ESO 341--IG04  are
shown in Figure \ref{fig_eso341}.  
\\

\end{document}